%% file: main.tex
\RequirePackage{fix-cm}
\documentclass[12pt,letterpaper]{article}
\usepackage[T1]{fontenc}
\usepackage[utf8]{inputenc}
\usepackage{libertine}
\usepackage[protrusion=false,expansion=false]{microtype}
\DisableLigatures{encoding=T1,shape=sc}
\usepackage[letterpaper,margin=1in]{geometry}
\usepackage{amsmath,amssymb,amsthm,mathtools,booktabs,array,tabularx,longtable,enumitem,needspace}
\usepackage[table,dvipsnames]{xcolor}
\usepackage{caption}
\definecolor{PaperBlue}{HTML}{214D69}
\definecolor{PaperTeal}{HTML}{EAF4F2}
\definecolor{PaperGray}{HTML}{F3F5F7}
\usepackage{placeins}
\usepackage{tikz}
\usetikzlibrary{arrows.meta,positioning,calc}
\usepackage[hidelinks,bookmarksnumbered=true]{hyperref}
\hypersetup{pdfauthor={Zekun Ye},pdftitle={Separations Between Randomized and Exact Quantum Query Complexity for Boolean Functions}}
\allowdisplaybreaks[2]
\renewenvironment{abstract}{\begin{center}\bfseries\abstractname\end{center}\begin{quotation}\normalsize}{\end{quotation}}
\setlist{itemsep=3pt,topsep=5pt}
\AtBeginDocument{%
  \setlength{\abovedisplayskip}{9pt plus 2pt minus 4pt}%
  \setlength{\belowdisplayskip}{9pt plus 2pt minus 4pt}%
  \setlength{\abovedisplayshortskip}{0pt plus 2pt}%
  \setlength{\belowdisplayshortskip}{5pt plus 2pt minus 2pt}%
}
\makeatletter
\renewcommand\footnotesize{\@setfontsize\footnotesize{11pt}{13.2pt}}
\renewcommand{\section}{\@startsection{section}{1}{\z@}
  {-3.5ex \@plus -1ex \@minus -.2ex}{2.3ex \@plus .2ex}
  {\normalfont\Large\scshape\bfseries\raggedright}}
\renewcommand{\subsection}{\@startsection{subsection}{2}{\z@}
  {-3.25ex \@plus -1ex \@minus -.2ex}{1.5ex \@plus .2ex}
  {\normalfont\large\scshape\bfseries\raggedright}}
\renewcommand{\subsubsection}{\@startsection{subsubsection}{3}{\z@}
  {-3.25ex \@plus -1ex \@minus -.2ex}{1.5ex \@plus .2ex}
  {\normalfont\normalsize\scshape\bfseries\raggedright}}
\renewcommand{\paragraph}{\@startsection{paragraph}{4}{\z@}
  {2.25ex \@plus 1ex \@minus .2ex}{-1em}
  {\normalfont\normalsize\bfseries}}
\makeatother
\numberwithin{equation}{section}
\newtheorem{theorem}{Theorem}[section]
\newtheorem{lemma}[theorem]{Lemma}
\newtheorem{proposition}[theorem]{Proposition}
\newtheorem{corollary}[theorem]{Corollary}
\theoremstyle{definition}\newtheorem{definition}[theorem]{Definition}

\newcommand{\E}{\mathbb E}
\newcommand{\R}{\operatorname R}\newcommand{\QE}{\operatorname Q_E}
\newcommand{\Rcc}{\operatorname R^{\mathrm{pub},\mathrm{cc}}}
\newcommand{\Qecc}{\operatorname Q_E^{\mathrm{cc}}}\newcommand{\Qeccs}{\operatorname Q_E^{\mathrm{cc},*}}
\newcommand{\AND}{\operatorname{AND}}\newcommand{\IP}{\operatorname{IP}}
\newcommand{\rank}{\operatorname{rank}}\newcommand{\dom}{\operatorname{dom}}

\newcommand{\TV}{d_{\mathrm{TV}}}

\begin{document}
\hyphenation{OpenAI}
\input{sections/frontmatter}
\input{sections/overview}
\input{sections/preliminaries}

\input{sections/total}
\input{sections/constant}
\appendix
\input{sections/communication_appendix}
\input{sections/acknowledgments}
\input{sections/ai_methodology}

\input{sections/references}
\end{document}

%% file: sections/frontmatter.tex
\hypersetup{pageanchor=false}
\begin{titlepage}
\begin{center}
{\LARGE\bfseries Separations Between Randomized and\\
Exact Quantum Query Complexity\\
for Boolean Functions\par}
\vspace{0.8em}
\input{authors}
\end{center}
\vspace{0.4em}
\begin{abstract}
We construct a family of total Boolean functions $F_{\mathrm{tot}}$ whose randomized and exact quantum query complexities satisfy
$R(F_{\mathrm{tot}})\ge Q_E(F_{\mathrm{tot}})^{2-o(1)}$.
Chiruvella, Kayal, Mittal, Paraashar, and Saurabh (\href{https://arxiv.org/abs/2611.16412v1}{2026}) also obtained a nearly quadratic separation for total functions in the query model by a different method.

For every fixed $\varepsilon\in(0,1)$, we construct a family of partial Boolean functions $F_{\mathrm{par}}$ on $n$ bits satisfying
$Q_E(F_{\mathrm{par}})=O_\varepsilon(1)$ and $R(F_{\mathrm{par}})=\Omega_\varepsilon(n^{1-\varepsilon})$.
This matches the asymptotic separation previously obtained with bounded error (\href{https://doi.org/10.1145/3406325.3451040}{STOC~2021}; \href{https://doi.org/10.1137/22M1468943}{SICOMP~2023}).
Lifting yields a family of partial communication functions $G_{\mathrm{par}}$ with
$\Qecc(G_{\mathrm{par}})=O_\varepsilon(\log n)$ and $\Rcc(G_{\mathrm{par}})=\Omega_\varepsilon(n^{1-\varepsilon})$, where $n$ is the total number of input bits held by the two parties.
\end{abstract}
\thispagestyle{empty}
\end{titlepage}
\hypersetup{pageanchor=true}
\pagenumbering{roman}
\begingroup
\setlength{\parskip}{0pt}
\tableofcontents
\par\endgroup
\clearpage
\pagenumbering{arabic}

%% file: authors.tex
{\large Zekun Ye\par}
\vspace{0.4em}
{\normalsize College of Computer and Data Science\par}
{\normalsize Fuzhou University, Fuzhou 350108, China\par}
{\normalsize\href{mailto:yezekun@fzu.edu.cn}{\texttt{yezekun@fzu.edu.cn}}\par}

%% file: sections/overview.tex
\section{Introduction}\label{sec:overview}

How much can quantum computation outperform randomized computation when the quantum algorithm must always be correct?
The query model makes this question accessible to unconditional upper and lower bounds: an algorithm accesses an input $x\in\{0,1\}^n$ by querying its bits, and only the number of queries is charged.
We consider Boolean functions $f:\dom(f)\subseteq\{0,1\}^n\to\{0,1\}$.
The function is \emph{total} if $\dom(f)=\{0,1\}^n$ and \emph{partial} if correctness is required only on a specified promise domain.

The \emph{deterministic query complexity} $D(f)$ is the minimum number of queries required in the worst case by a deterministic algorithm that computes $f$.
The \emph{randomized query complexity} $R(f)$ and the \emph{quantum query complexity} $Q(f)$ allow error at most $1/3$ on every input in $\dom(f)$.
The \emph{exact quantum query complexity} $Q_E(f)$ requires the quantum algorithm to output $f(x)$ with probability one on every such input.
In each case, the query bound holds on every valid input and every computational branch; quantum queries may access input positions in superposition.
Section~\ref{sec:preliminaries} gives the formal definitions.

Comparing $R(f)$ with $Q_E(f)$ asks how much quantum computation can outperform randomized computation when the classical algorithm may have bounded error but the quantum algorithm must always be correct.
Since $Q(f)\le Q_E(f)$, a large gap between $R(f)$ and $Q_E(f)$ demonstrates a quantum advantage that persists under this stronger correctness requirement.

The total and partial settings pose different quantitative questions.
For every total Boolean function, the general bound
$R(f)\le D(f)\le16Q_E(f)^3$~\cite{Mid04} limits the separation to a polynomial.
Chiruvella, Kayal, Mittal, Paraashar, and Saurabh~\cite{CKMPS26} recently obtained a nearly quadratic separation for total functions.
For partial functions, quantum algorithms with bounded error can use $O_\varepsilon(1)$ queries while randomized algorithms need $\Omega_\varepsilon(n^{1-\varepsilon})$ queries for every fixed $\varepsilon\in(0,1)$~\cite{BS21,SSW23}.
Girish and Servedio~\cite[Section~1.3]{GS26} explicitly asked whether a comparable separation can be attained with exact quantum algorithms, and proposed extremal $k$-Forrelation as a candidate.
We obtain a nearly quadratic separation for total functions and construct partial functions with constant exact quantum query complexity and nearly linear randomized query complexity.

We also study the corresponding question in communication complexity.
Here Alice receives $x\in\mathcal X$ and Bob receives $y\in\mathcal Y$; they exchange messages to compute a Boolean function $G(x,y)$.
The cost is the total number of bits or qubits exchanged; local computation is unrestricted.
The function is total if it is defined on every pair in $\mathcal X\times\mathcal Y$, and partial if the pair is restricted by a promise.
Randomized protocols exchange bits, whereas quantum protocols may exchange qubits.
Our query constructions yield corresponding communication separations through randomized lifting and exact quantum simulation.

\subsection{Main Results}

We write $\Rcc$ for randomized communication complexity with public randomness, $\Qecc$ for exact quantum communication complexity without prior entanglement, and $\Qeccs$ for exact quantum communication complexity with prior entanglement.
Randomized algorithms and protocols have error at most $1/3$.
Table~\ref{tab:main-results} summarizes our results.
The query separations are proved in Sections~\ref{sec:total} and~\ref{cp-section}; Appendix~\ref{app:communication} derives their communication consequences.
All bounds hold along the constructed families as the input length grows, and an $\varepsilon$ subscript allows constants to depend on the fixed $\varepsilon\in(0,1)$.

\begin{table}[htbp]
\centering
\caption{Our separations in query and communication complexity. Here $n$ is the total input length, and $q$ is the exact quantum complexity of the function in the corresponding row.}
\label{tab:main-results}
\renewcommand{\arraystretch}{1.2}
\begin{tabularx}{\textwidth}{@{}>{\raggedright\arraybackslash}p{0.27\textwidth}>{\raggedright\arraybackslash}p{0.25\textwidth}>{\raggedright\arraybackslash}X@{}}
\toprule
Model and function type & Exact quantum complexity & Randomized lower bound\\
\midrule
Query: total & $q$ & $q^{2-o(1)}$\\
Query: partial & $O_\varepsilon(1)$ & $\Omega_\varepsilon(n^{1-\varepsilon})$\\
Communication: total & $q$ & $q^{2-o(1)}$\\
Communication: partial & $O_\varepsilon(\log n)$ & $\Omega_\varepsilon(n^{1-\varepsilon}(\log n)^\varepsilon)$\\
\bottomrule
\end{tabularx}
\end{table}

\begin{theorem}[Total functions]\label{thm:main-total}
There is a family of total Boolean functions $F_{\mathrm{tot}}$ such that, writing $q=Q_E(F_{\mathrm{tot}})$,
\[
 R(F_{\mathrm{tot}})\ge q^2\,2^{-O(\sqrt{\log q})}.
\]
The input length is $q^{O(1)}$.
\end{theorem}

Lifting this family gives total communication functions $G$ with $\Rcc(G)\ge q^2\,2^{-O(\sqrt{\log q})}$ and total input length $q^{O(1)}$, for either $q=\Qecc(G)$ or $q=\Qeccs(G)$; see Corollary~\ref{thm:total-communication} in Appendix~\ref{app:total-communication}.

\begin{theorem}[Partial functions]\label{thm:main-partial}
For every fixed $\varepsilon\in(0,1)$ and every sufficiently large $n$, there is a partial Boolean function $F_{\mathrm{par}}$ on $n$ bits satisfying
\[
 Q_E(F_{\mathrm{par}})=O_\varepsilon(1),
 \qquad R(F_{\mathrm{par}})=\Omega_\varepsilon(n^{1-\varepsilon}).
\]
\end{theorem}

Randomized lifting~\cite{CFKMP19} and exact quantum query simulation~\cite{BCW98} give the following corollary.

\begin{corollary}[Partial functions]\label{cor:main-communication}
For every fixed $\varepsilon\in(0,1)$ and every sufficiently large $n$, there is a partial communication function $G_{\mathrm{par}}$ satisfying
\[
 \Qecc(G_{\mathrm{par}})=O_\varepsilon(\log n),
 \qquad \Rcc(G_{\mathrm{par}})=\Omega_\varepsilon\!\left(n^{1-\varepsilon}(\log n)^\varepsilon\right),
\]
where $n$ is the total number of input bits held by the two parties.
\end{corollary}

\FloatBarrier

\subsection{Related Work}\label{sec:related-work}

We review earlier separations for total functions and partial functions in the query model, followed by results in communication complexity.

\paragraph{Total functions in the query model.}
Early exact quantum algorithms for total Boolean functions provided mainly speedups by a constant factor over classical algorithms.
Ambainis~\cite{Amb16} gave the first superlinear separation using an iterated function: on $N$ input bits, its exact quantum query complexity is $O(N^{0.8675\ldots})$, whereas its randomized query complexity is $\Omega(N)$.

Ambainis, Balodis, Belovs, Lee, Santha, and Smotrovs~\cite{ABB17} subsequently used pointer functions to obtain
$R(f)=\widetilde\Omega(Q_E(f)^{3/2})$.

The cheat sheet framework of Aaronson, Ben-David, and Kothari~\cite{ABK16} converts partial functions into total functions by appending a certificate array.
The answers to several instances of the original function specify a certificate location, and the total function checks the validity of that certificate.
The exact quantum upper bound depends on the exact cost of computing the original function and verifying the certificate~\cite[Theorem~16]{ABK16}.

Recent work has also studied exact quantum algorithms for compositions.
Pednekar and Podder~\cite[Section~5]{PP26} showed that exact quantum query complexity need not compose multiplicatively.

Chiruvella, Kayal, Mittal, Paraashar, and Saurabh~\cite[Theorems~9.3 and~11.5]{CKMPS26} obtained nearly quadratic separations for total functions in the query and communication models using exact algorithms for recursively composed functions.
Our construction uses pointers and verifies the original address to determine and rescale the acceptance probability, using the method of~\cite[Section~7]{AIK26}.
Our bounds have a smaller subpolynomial loss: $q^2 2^{-O(\sqrt{\log q})}$ in Theorem~\ref{thm:main-total} and Corollary~\ref{thm:total-communication}, compared with $q^2\exp(-O(\sqrt{\log q\,\log\log q}))$ in~\cite{CKMPS26}, where $q$ denotes the exact quantum complexity in the corresponding model.
The nearly quadratic communication separations in both papers follow from randomized lifting with the inner product gadget and exact quantum query simulation.

\paragraph{Partial functions in the query model.}
Simon's problem gives an exponential quantum query advantage over randomized algorithms~\cite{Sim97}.
Brassard and H{\o}yer~\cite{BH97} developed an exact quantum algorithm for Simon's problem using a polynomial number of queries in the worst case.

Forrelation and its variants have progressively increased the separation between randomized and quantum query complexity with bounded error for partial functions.
Forrelation measures the correlation of one Boolean function with the Fourier transform of another.
Aaronson and Ambainis~\cite{AA18} proved that its quantum query complexity is $O(1)$ and its randomized query complexity is $\Omega(\sqrt N/\log N)$, where $N$ denotes the input length.
They also introduced $k$-Forrelation to study randomized lower bounds as a function of quantum query complexity.

For the variant $k$-Rorrelation, Tal~\cite{Tal20} obtained a separation of $O_\varepsilon(1)$ quantum queries from $\Omega_\varepsilon(N^{2/3-\varepsilon})$ randomized queries for every fixed $\varepsilon>0$.
Bansal and Sinha~\cite{BS21} then proved a randomized lower bound for $k$-Forrelation, obtaining $O_\varepsilon(1)$ quantum queries versus $\Omega_\varepsilon(N^{1-\varepsilon})$ randomized queries for every fixed $\varepsilon\in(0,1)$.
Sherstov, Storozhenko, and Wu~\cite{SSW23} independently obtained the same asymptotic separation for $k$-Rorrelation.

In the exact quantum model, Girish and Servedio~\cite{GS26} studied extremal Forrelation and proved a separation of one exact quantum query from $\widetilde\Omega(N^{1/4})$ randomized queries.
Canonne, Chen, and Mestre~\cite{CCM26} subsequently proved an $N^{1/2-o(1)}$ lower bound for \emph{nonadaptive} randomized algorithms for extremal Forrelation.

Exact quantum algorithms for promise problems also appear in recent oracle separations.
Mantri~\cite[Theorem~5.1]{Man26} evaluates a recursive addressing problem of height $h$ using $3^h$ exact quantum queries.
Oral and Podder~\cite[Proposition~3.2]{OP26} give an exact algorithm using three queries for Forrelation promised to equal $1/2$ or $-1/2$.
For this promise, the estimation algorithm of Bravyi, Gosset, Grier, and Schaeffer~\cite[Theorem~1]{BGGS21} gives randomized query complexity $O(\sqrt N)$, where $N$ is the total length of the two truth tables.

Our exact quantum query complexity is bounded by a constant depending on $\varepsilon$, and our randomized lower bounds hold for adaptive algorithms.
We obtain the separation discussed in~\cite[Section~1.3]{GS26} for a new partial function.

\paragraph{Communication complexity.}
For total functions, Anshu, Belovs, Ben-David, G\"o\"os, Jain, Kothari, Lee, and Santha~\cite{ABB16} obtained a separation $\Rcc(G)=\widetilde\Omega(\Qecc(G)^{3/2})$.
They developed a communication version of the cheat sheet method and used information complexity to prove a classical lower bound allowing interaction and public randomness.
\cite[Theorem~11.5]{CKMPS26} subsequently obtained a nearly quadratic separation in this model.

When quantum protocols may have bounded error, recent work has obtained exponential quantum advantages for total functions in communication complexity.
Gavinsky~\cite{Gav26} first separated polylogarithmic quantum communication from polynomial randomized communication.
Le Gall~\cite{LG26} improved the randomized lower bound to $\widetilde\Omega(\sqrt n)$ with two quantum messages, and to $\Omega_\varepsilon(n^{1-\varepsilon})$ with a constant number of quantum messages for every fixed $\varepsilon\in(0,1)$, while retaining polylogarithmic quantum communication.
These quantum protocols use no prior entanglement, and the randomized lower bounds allow public randomness and arbitrarily many rounds.

For partial functions, Vector-in-Subspace in $\mathbb R^d$ admits an exact quantum protocol using $O(\log d)$ qubits, and Wang~\cite{Wang26} proved that its randomized communication complexity is $\Theta(\sqrt d)$.
Here $d$ is the dimension of the ambient space, and the inputs are a real vector and a subspace.

For functions on finite bit strings, lifting the extremal Forrelation separation of Girish and Servedio~\cite{GS26} gives $O(\log n)$ exact quantum communication complexity and $\widetilde\Omega(n^{1/4})$ randomized communication complexity.
We retain $O_\varepsilon(\log n)$ exact quantum communication complexity and increase the randomized lower bound to $\Omega_\varepsilon(n^{1-\varepsilon})$.

With bounded error on the quantum side, logarithmic quantum communication versus nearly linear randomized communication is already known for partial functions.
Bansal and Sinha~\cite[Corollary~1.6(a)]{BS21} and Sherstov, Storozhenko, and Wu~\cite[Theorem~1.6]{SSW23} each proved $O_\varepsilon(\log n)$ quantum communication versus $\Omega_\varepsilon(n^{1-\varepsilon})$ randomized communication.
We achieve this asymptotic separation with exact quantum protocols.

\subsection{Proof Overview}

We first construct total functions and partial functions that retain randomized hardness while admitting exact quantum algorithms, and then lift both separations to communication complexity.
For total functions, we combine the pointer construction and coupling argument of~\cite{OAI26} with the method of~\cite[Section~7]{AIK26} for rescaling the success probability.
Our adaptation verifies the original address in full, so every accepting branch contains the information needed to determine the acceptance probability.
For partial functions, we encode both an address and the exact success probability of its recovery circuit, and prove that the resulting encoding promise preserves randomized hardness.

\paragraph{Total functions.}
The key step is to transfer the hardness of the partial function $k$-Forrelation to a total Boolean function while preserving an efficient exact quantum algorithm.
We use the answers to several independent Forrelation instances as an address label and adapt the pointer construction with addresses of OpenAI~\cite[Section~2.2]{OAI26}.
The input has $m$ columns, each containing data bits, an address instance, and a pointer table.
For a valid address, a label function $P$ selects the true pointer cell.
A column is a \emph{winner} if its data bits are all one, its address is valid, and its true cell points to a zero in every other column; see Figure~\ref{fig:winner-construction}.
There is at most one winner, and the function records whether a winner exists.
A column whose address violates the Forrelation promise cannot be a winner; every other column is evaluated by the same conditions.
The function is therefore defined on every input string, including inputs whose addresses violate the promise.

\input{sections/winner_figure}

Each Forrelation instance used in an address has table length $m$ and order $\Theta(\sqrt{\log m})$, and $\Theta(\log m)$ answers form one label.
The bound in Section~\ref{tot:sec:main} gives a quantum query complexity of $m^{o(1)}$ for recovering each label, so recovering all labels uses $m^{1+o(1)}$ queries.
The original address in each column has length $m^{1+o(1)}$, and we set the data length to $b=m^{1+o(1)}$.
After decoding, the algorithm chooses a candidate through $m-1$ pairwise comparisons.
Each comparison reads only two pointers and their target bits, for a total of $O(m\log m)$ queries.
If the winner's label is decoded correctly, its selected pointer cell points to a zero in every opponent's column, whereas no pointer can point to a zero in the winner's column.
The winner therefore survives every comparison, without requiring any other label to be correct.

The algorithm then reads the candidate's entire original address and data block and checks the winner conditions, at a cost of $m^{1+o(1)}$ queries.
This full verification is performed only once, for the final candidate, and replaces the verification of circuit evaluations in~\cite{OAI26}.
Reading the original address determines its validity and true label.
Verification also requires the candidate's decoded label to be correct, making acceptance equivalent to correct decoding of the winner's own address.
The winner's original address thus determines the acceptance probability.
Using the weighting method of~\cite[Section~7]{AIK26}, we rescale this probability to $1/4$ on positive inputs and keep it zero on negative inputs.
One round of amplitude amplification~\cite{BHMT02} then computes the same total function exactly.
This adds only a constant factor to the query count, so decoding, comparison, and one full verification together give $Q_E(F_{\mathrm{tot}})\le m^{1+o(1)}$.

The randomized lower bound adapts the coupling argument of~\cite[Proposition~4.2]{OAI26}.
The Forrelation lower bound~\cite{BS21} and Lemma~\ref{lem:local-budget-product} give a hard address distribution under which any randomized algorithm making at most $R_{\mathrm{addr}}=m^{1-o(1)}$ queries predicts the label with probability at most $m^{-5}$.
In a negative input, the column addresses are independent samples from this distribution, each data column has one independently and uniformly hidden zero, and the pointer tables are fixed.
We independently choose a uniform column, delete its zero, and modify its true pointer cell to make it a winner.
The two inputs differ only at these locations.
When run with the same random coins, the algorithm receives identical answers until it queries one of them.

Consider an algorithm making at most $T$ queries, and follow its execution on the negative input.
At most $2T/b$ columns receive queries to more than half their data positions.
In any column whose zero has not been found and with at most half its positions queried, a fresh data query finds the zero with conditional probability at most $2/b$.
The expected number of columns whose zero is found is therefore at most $4T/b$.
Since the winner's column is uniform and independent of this execution, the probability of hitting the deleted zero is at most $4T/(mb)$.
Likewise, at most $T/R_{\mathrm{addr}}$ columns receive more than $R_{\mathrm{addr}}$ address queries, so the probability that the winner lies among them is at most $T/(mR_{\mathrm{addr}})$.
If the algorithm queries the true pointer cell after at most $R_{\mathrm{addr}}$ address queries, its behavior gives a procedure for predicting the label with the same query bound.
Averaging over the hidden column and taking a union bound over $T$ steps bounds this event by $Tm^{-5}$.
A randomized algorithm with error at most $1/3$ has acceptance probabilities differing by at least $1/3$ on the two input distributions, so
\[
 \frac13\le\frac{4T}{mb}+\frac{T}{mR_{\mathrm{addr}}}+Tm^{-5}.
\]
Substituting $b=m^{1+o(1)}$ and $R_{\mathrm{addr}}=m^{1-o(1)}$ gives $T\ge m^{2-o(1)}$.
Thus $R(F_{\mathrm{tot}})\ge m^{2-o(1)}$, yielding a nearly quadratic separation from the exact quantum upper bound above.

\FloatBarrier
\paragraph{Partial functions.}
For partial functions, we use two encodings to make the numbers of queries needed to recover the address and its success probability independent of the input length.
Fix $\varepsilon\in(0,1)$, and choose the Forrelation order $k$ and the number $d$ of address blocks depending only on $\varepsilon$; each Forrelation table has length $M$.
The base input consists of $J=dM^{1/d}$ masked $k$-Forrelation instances, promised to have an answer string equal to a blockwise Hadamard encoding of an address of $\log M$ bits.
The quantum algorithm runs the Forrelation algorithm in a superposition of instance indices and performs Bernstein--Vazirani decoding on each block.
Each block uses a constant number of Forrelation circuit calls, whose query cost depends only on $k,d$.
The address is therefore recovered with probability at least $24/25$ using $O_{k,d}(1)$ queries.

The address algorithm is fixed in advance and queries only the base input.
Its gates are Hadamard gates, queries, and signed permutations, so its success probability has a finite binary representation of length $O_{k,d}(\log M)$.
We divide this representation into blocks of length $(\log M)/d$, encode each block using the Hadamard code, and store the encodings in the true address cell of an auxiliary table.
On negative inputs, the table is all zero.
On positive inputs, only the true cell contains a flag equal to one and the correct probability encoding; all other cells are zero.
These conditions define the promise of the new partial function, with the success probability determined solely by the base input.
After recovering a candidate address, the algorithm reads the flag and coherently decodes the probability.
Whenever the flag is one, the probability is correct, so the algorithm can rescale the acceptance probability to $1/4$ and apply one round of amplitude amplification.
The number of probability blocks depends only on $k,d$, and each block is decoded exactly with one query.
Address recovery, probability retrieval, and amplitude amplification thus use a number of queries depending only on $\varepsilon$:
\[
 Q_E(F_{\mathrm{par}})=O_\varepsilon(1).
\]

The randomized lower bound must show that the probability table does not make the true address easy to find.
We first choose a hard distribution for independent Forrelation instances and then condition on the encoding promise.
Reducing the prediction advantage for each answer to $O(1/J)$ ensures that the full address can be predicted with probability only $O(1/M)$ when the algorithm makes at most $cR(\operatorname{Forr}_{k,M})/J^2$ queries, for a sufficiently small absolute constant $c>0$.
We then follow the algorithm's queries when the table is zero and bound the first visit to the true cell; before this visit, the two input types give identical answers.
Combined with the Forrelation lower bound~\cite{BS21}, this gives a randomized lower bound of $\Omega_{k,d}(M^{1-1/k-2/d}/\log M)$.
Including the base input and all probability encodings, the total input length is $n=\Theta_{k,d}(M^{1+1/d})$.
Taking $k=\lceil4/\varepsilon\rceil$ and $d=\lceil12/\varepsilon\rceil$ gives $R(F_{\mathrm{par}})=\Omega_\varepsilon(n^{1-\varepsilon/2}/\log n)$, and the remaining slack absorbs the logarithmic factor.
Consequently,
\[
 R(F_{\mathrm{par}})=\Omega_\varepsilon(n^{1-\varepsilon}).
\]

\paragraph{Communication lifting.}
We replace each input bit by an inner product gadget of logarithmic length.
The randomized lifting theorem~\cite{CFKMP19} transfers query lower bounds to randomized communication with public randomness, while exact query simulation~\cite{BCW98} replaces each quantum query by logarithmic communication.
For partial functions, accounting for the input length gives logarithmic exact quantum communication complexity and nearly linear randomized communication complexity.

\subsection{Organization}

Section~\ref{sec:preliminaries} introduces the models and tools.
Section~\ref{sec:total} proves the totalization theorem and uses Forrelation addresses to obtain the nearly quadratic separation for total functions.
Section~\ref{cp-section} gives the construction and separation for partial functions.
Appendix~\ref{app:communication} derives the communication separations from these query results.

%% file: sections/winner_figure.tex
\begin{figure}[!htbp]
\centering
\begin{tikzpicture}[x=1cm,y=1cm,>=Latex,font=\normalsize,
  every node/.style={text=black!85},
  data/.style={draw=black!50,line width=.55pt,fill=black!2,minimum width=3.6cm,minimum height=.64cm,inner sep=2pt},
  address/.style={draw=black!50,line width=.55pt,fill=black!2,rounded corners=2pt,minimum width=3.6cm,minimum height=.65cm,inner sep=2pt},
  ptr/.style={draw=black!50,line width=.55pt,fill=white,minimum width=3.6cm,minimum height=.62cm,inner sep=2pt},
  active/.style={ptr,draw=blue!45!black,fill=teal!10,line width=1.1pt}]
\begin{scope}[shift={(0,0)}]
\node at (0,4.85) {Column $i$};
\node[data] at (0,4.0) {};
\node at (-1.28,4.0) {$1$};\node at (-.66,4.0) {$\cdots$};
\node[draw=blue!45!black,fill=teal!8,circle,line width=.65pt,inner sep=1.5pt] (zeroi) at (0,4.0) {$0$};
\node at (.66,4.0) {$\cdots$};\node at (1.28,4.0) {$1$};
\node[address] at (0,2.45) {$x_i\in\{0,1\}^a$};
\node at (0,1.3) {$K$ cells};
\draw[draw=black!50,line width=.55pt] (-1.8,-.82) rectangle (1.8,.80);
\node at (0,.57) {$\vdots$};\node at (0,-.57) {$\vdots$};
\node[ptr] at (0,0) {$\pi_i(z,\cdot)$};
\end{scope}
\begin{scope}[shift={(4.8,0)}]
\node at (0,4.85) {Winner $w$};
\node[data,draw=blue!45!black,fill=teal!10] at (0,4.0) {};
\node at (0,4.0) {$y_w=1^b$};
\node[address] at (0,2.45) {$x_w\in\dom(P)$};
\node at (0,1.3) {$K$ cells};
\draw[draw=black!50,line width=.55pt] (-1.8,-.82) rectangle (1.8,.80);
\node at (0,.57) {$\vdots$};\node at (0,-.57) {$\vdots$};
\node[active] (winnercell) at (0,0) {$z^\star=P(x_w)$};
\end{scope}
\begin{scope}[shift={(9.6,0)}]
\node at (0,4.85) {Column $j$};
\node[data] at (0,4.0) {};
\node at (-1.28,4.0) {$1$};\node at (-.66,4.0) {$\cdots$};
\node[draw=blue!45!black,fill=teal!8,circle,line width=.65pt,inner sep=1.5pt] (zeroj) at (0,4.0) {$0$};
\node at (.66,4.0) {$\cdots$};\node at (1.28,4.0) {$1$};
\node[address] at (0,2.45) {$x_j\in\{0,1\}^a$};
\node at (0,1.3) {$K$ cells};
\draw[draw=black!50,line width=.55pt] (-1.8,-.82) rectangle (1.8,.80);
\node at (0,.57) {$\vdots$};\node at (0,-.57) {$\vdots$};
\node[ptr] at (0,0) {$\pi_j(z,\cdot)$};
\end{scope}
\draw[->,line width=1.05pt,blue!45!black] (winnercell.west) -- (2.35,0) -- (2.35,3.2) -- (0,3.2) -- (zeroi.south);
\draw[->,line width=1.05pt,blue!45!black] (winnercell.east) -- (7.25,0) -- (7.25,3.2) -- (9.6,3.2) -- (zeroj.south);
\end{tikzpicture}
\caption{The winner and its true pointer cell. Each of the $m$ columns contains, from top to bottom, $b$ data bits, an original address of $a$ bits, and a table of $K$ pointer cells. Each cell stores a pointer to every other column. The winner $w$ has data $y_w=1^b$, a valid address, and a true cell $z^\star=P(x_w)$ pointing to a zero in every other column. Only two other columns are shown; the unmarked data bits are illustrative.}
\label{fig:winner-construction}
\end{figure}
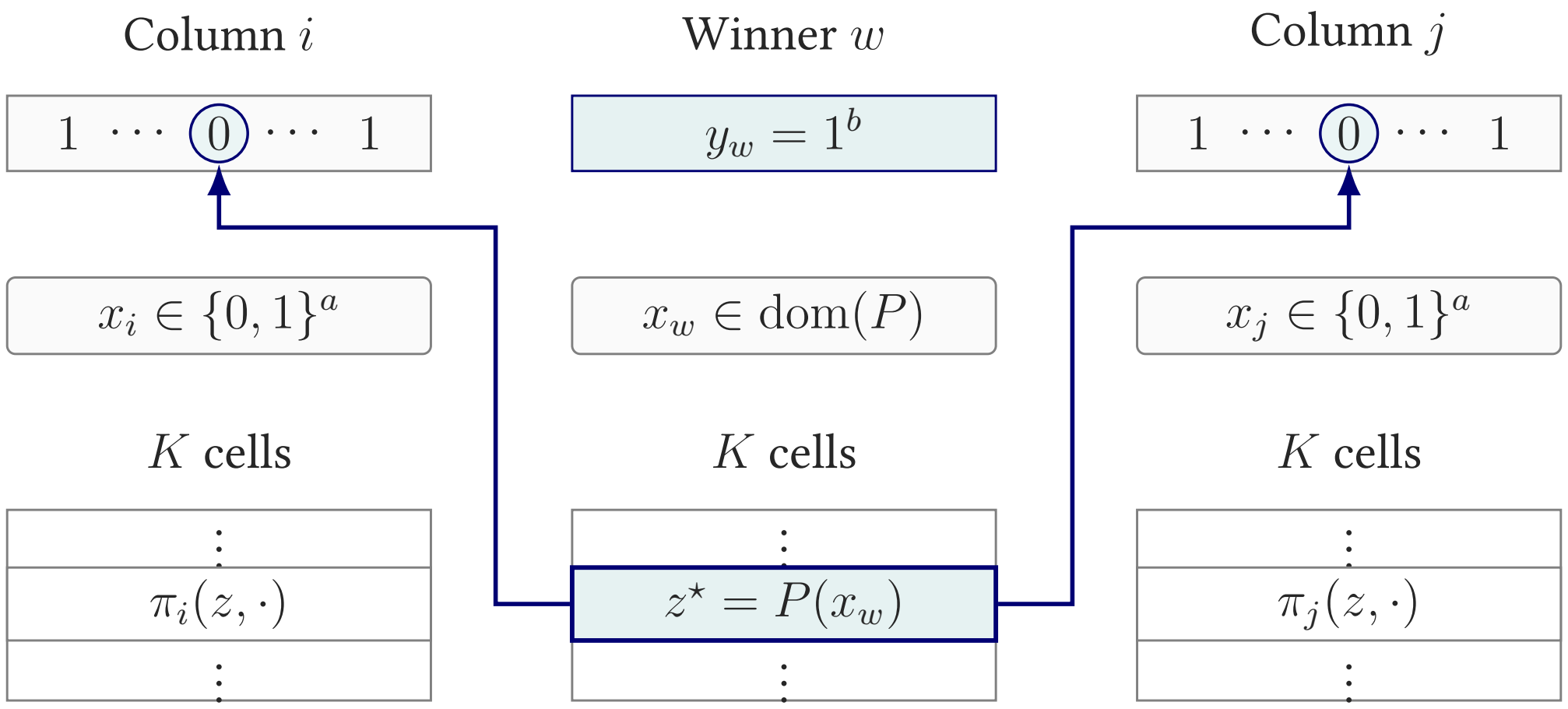

%% file: sections/preliminaries.tex
\section{Preliminaries}\label{sec:preliminaries}

We define the query and communication models, then state the restriction, decoding, amplitude amplification, and direct product tools used in the proofs.

\subsection{Computational Models and Notation}

Write $[m]=\{1,\ldots,m\}$. A partial Boolean function is a map
$f:\dom(f)\to\{0,1\}$ with promise domain $\dom(f)\subseteq\{0,1\}^n$;
it is total if $\dom(f)=\{0,1\}^n$.
All correctness requirements below apply to every input in the promise domain.

\paragraph{Query complexity.}
An algorithm accesses an unknown input $x\in\dom(f)$ through queries.
A classical query specifies a coordinate $i\in[n]$ and returns $x_i$.
Each query may depend on the previous answers and, for a randomized algorithm, on its internal random coins.
The \emph{deterministic query complexity} $D(f)$ is the minimum number of queries needed by a deterministic algorithm that computes $f$ without error.
The \emph{randomized query complexity} $R(f)$ is defined similarly, allowing internal randomness and error probability at most $1/3$ on every input in $\dom(f)$.
The number of queries is bounded on every valid input and for every outcome of the random coins.

A quantum algorithm uses the standard bit oracle~\cite[Section~3.2]{BCDMW01}
\[
 O_x|i,b,z\rangle=|i,b\oplus x_i,z\rangle,
\]
where $i$ is the query index, $b$ is the answer bit, and $z$ is the workspace.
The oracle acts linearly on superpositions of query indices.
An algorithm making $T$ queries starts in a fixed state and alternates oracle calls with unitaries independent of the input:
\[
 |\psi_x\rangle=U_T O_x U_{T-1}\cdots O_x U_0|0\rangle.
\]
It then measures a designated output bit.
The \emph{quantum query complexity} $Q(f)$ is the minimum number of queries needed to compute $f$ with error probability at most $1/3$ on every input in $\dom(f)$.
The \emph{exact quantum query complexity} $Q_E(f)$ is defined in the same way, but the output must equal $f(x)$ with probability one on every such input.
Here the probability is determined by the output measurement.

Only oracle calls contribute to query complexity; operations between queries are not counted.
We allow arbitrary fixed exact unitaries.
Intermediate measurements can be recorded coherently in workspace registers.
Retaining these records lets us use a unitary circuit and its inverse.

\paragraph{Communication complexity.}
Let $G:\dom(G)\subseteq\mathcal X\times\mathcal Y\to\{0,1\}$, where $\mathcal X$ and $\mathcal Y$ are finite sets of binary strings.
Alice receives $x\in\mathcal X$ and Bob receives $y\in\mathcal Y$, with the promise that $(x,y)\in\dom(G)$.
They may interact for any number of rounds, and Bob outputs the answer.
The function is total when $\dom(G)=\mathcal X\times\mathcal Y$.
The communication cost is the total number of bits or qubits exchanged by the parties; local computation is not counted~\cite[Section~1.1]{BCW98}.

In a randomized protocol, each message depends on the sender's input, the messages exchanged so far, and the random coins.
Public randomness is a shared random string independent of the inputs; private randomness is also allowed.
The \emph{randomized communication complexity with public randomness} $\Rcc(G)$ is the minimum integer $c$ for which such a protocol $\Pi$ communicates at most $c$ bits on every valid input pair and every outcome of the random coins, and satisfies
\[
 \Pr[\Pi(x,y)=G(x,y)]\ge\frac23
 \qquad\text{for every }(x,y)\in\dom(G).
\]
The probability is over all public and private coins.

In a quantum protocol, the parties apply local quantum operations controlled by their respective inputs and exchange quantum registers; Bob finally measures an output bit.
The \emph{exact quantum communication complexity} $\Qecc(G)$ is the minimum integer $c$ for which a protocol without prior entanglement communicates at most $c$ qubits on every valid input pair and every possible measurement branch, and satisfies
\[
 \Pr[\Pi(x,y)=G(x,y)]=1
 \qquad\text{for every }(x,y)\in\dom(G).
\]
The \emph{exact quantum communication complexity with prior entanglement} $\Qeccs(G)$ has the same definition, except that the parties may share an arbitrary entangled state of finite dimension, independent of the inputs, before communication begins.
Local auxiliary registers do not contribute to communication cost.
Replacing success probability $1$ by at least $2/3$ gives the corresponding quantum communication model with bounded error.
All quantum communication upper bounds in this paper are achieved without prior entanglement.

\paragraph{Notation.}
The variable $n$ denotes the total input length of the current problem; in the communication model, it is the sum of the two parties' input lengths.
We use $n_0$ for the query input length before lifting.
All logarithms are base $2$, except for $\ln$.
An $\varepsilon$ subscript on asymptotic notation allows the hidden constants to depend on the fixed parameter $\varepsilon\in(0,1)$.
For probability distributions $\mu,\nu$, their total variation distance is
$\TV(\mu,\nu):=\sup_E|\mu(E)-\nu(E)|$, where the supremum ranges over all events.

We use $f$ for a general Boolean function and $\operatorname{Forr}_{k,N}$ for the $k$-Forrelation partial Boolean function with $N$ entries in each sign table.
Its associated correlation is denoted by $\Phi_{k,N}$.
Our total and partial functions in the query model are $F_{\mathrm{tot}}$ and $F_{\mathrm{par}}$, and their communication versions are $G_{\mathrm{tot}}$ and $G_{\mathrm{par}}$.
The general communication function in the lifting framework is denoted by $G$.
The parameters of each construction are collected in its notation table.

\paragraph{Research methodology.}
ChatGPT assisted with developing and organizing the constructions and proof arguments in Sections~\ref{sec:total}--\ref{cp-section} and Appendix~\ref{app:communication}, including the exact algorithms and proofs of the randomized lower bounds.
Its role is described in the AI Methodology Statement at the end of the paper.

\subsection{Basic Tools for Query Complexity}

Restrictions give exact quantum lower bounds for the total construction.
Bernstein--Vazirani decoding and amplitude amplification yield exact algorithms, whereas Yao's principle and a direct product bound yield randomized lower bounds.

\begin{lemma}[Exact query complexity of AND; consequence of {\cite[Proposition~6.1]{BCDMW01}}]\label{prelim:lem:and}
For every integer $b\ge1$, we have $Q_E(\AND_b)=b$.
\end{lemma}

A \emph{restriction} $g$ of $f$ is obtained by fixing some of its input bits.
An algorithm for $f$ can be simulated on $g$ without querying the fixed bits, so $Q_E(g)\le Q_E(f)$.
In particular, if fixing some input bits gives $\AND_b$ on all $b$ remaining bits, then
\begin{equation}\label{prelim:eq:and-restriction}
 Q_E(f)\ge Q_E(\AND_b)=b.
\end{equation}

\begin{lemma}[Yao's minimax principle~{\cite{Yao77}}]\label{prelim:lem:yao}
Let $g:D\to\Sigma$, where $D\subseteq\{0,1\}^m$ is nonempty and $\Sigma$ is finite.
Fix an integer $T\ge0$ and $p\in[0,1]$.
If every randomized algorithm using at most $T$ queries has an input on which its success probability is at most $p$, then there is a distribution $\mu$ on $D$ such that every randomized algorithm $A$ using at most $T$ queries satisfies
\[
 \Pr_{x\sim\mu,A}[A(x)=g(x)]\le p.
\]
\end{lemma}

\begin{lemma}[Bernstein--Vazirani decoding~{\cite[Section~3]{CEMM98}}]\label{prelim:lem:bv}
For $u\in\mathbb F_2^b$, define $\operatorname{Enc}(u)_r=u\cdot r$ for $r\in\mathbb F_2^b$.
There is a quantum algorithm that recovers $u$ exactly and coherently with one bit query to $\operatorname{Enc}(u)$, leaving its auxiliary registers in a fixed state independent of $u$.
\end{lemma}

\begin{lemma}[Exact amplitude amplification; special case of {\cite[Section~2]{BHMT02}}]\label{prelim:lem:exact-amplification}
Let $W$ be a unitary query algorithm using at most $T$ queries, whose accepting subspace is specified by a projector $\Pi$ independent of the input.
If
\[
 \|\Pi W_x|0\rangle\|^2=\frac{f(x)}4
 \qquad\text{for every }x\in\dom(f),
\]
then $f$ has an exact quantum algorithm using at most $3T$ queries.
\end{lemma}

This form follows directly from one round of amplitude amplification in~\cite[Equations~(1) and~(8)]{BHMT02}: when the accepting probability is $1/4$, one round increases it to $1$ and calls $W$ twice and $W^\dagger$ once.

The following direct product lemma is stated and proved in~\cite[Lemma~4.1]{OAI26}.
It permits adaptive interleaving while bounding the number of queries to each instance separately.

\begin{lemma}[Direct product lemma]\label{lem:local-budget-product}
Let $P:D\to\{0,1\}$, where $D\subseteq\{0,1\}^m$, and let $\nu$ be a distribution on $D$.
Let $L$ be a nonnegative integer.
Suppose that every randomized algorithm using at most $L$ queries computes $P$ with success probability at most $p_0$ under $\nu$.
Given $r$ independent samples from $\nu$, a randomized algorithm may interleave its queries adaptively but makes at most $L$ queries to each instance.
Then the probability that it outputs all $r$ answers correctly is at most $p_0^r$.
\end{lemma}

%% file: sections/total.tex
\section{Total Functions}\label{sec:total}\label{tot:sec:totalization}

We adapt the pointer construction of~\cite[Section~2.2]{OAI26} to obtain a total Boolean function.
The quantum algorithm reads and verifies the original address in full, obtaining the information needed to rescale its acceptance probability.
The randomized lower bound follows from a coupling that plants a winning column.
We first combine these bounds into a totalization theorem, then instantiate the addresses with $k$-Forrelation and choose the parameters to obtain a nearly quadratic separation.

Let $P:\dom(P)\subseteq\{0,1\}^{a}\to[K]$ be a label function with a nonempty promise domain.
Fix a quantum algorithm $\mathcal B$ that makes $Q_{\mathrm{addr}}$ queries and outputs the correct label with probability at least $2/3$ on every valid address.
The algorithm makes the same number of queries outside the promise domain and always outputs a label in $[K]$.

Let $\mu$ be supported on $\dom(P)$, and let $R_{\mathrm{addr}}\ge1$ be an integer such that every randomized algorithm $A$ making at most $R_{\mathrm{addr}}$ queries on every input satisfies
\begin{equation}\label{tot:eq:address-hardness}
 \Pr_{x\sim\mu}[A(x)=P(x)]\le p.
\end{equation}
The probability includes the internal randomness of $A$, and the algorithms may be fully adaptive.
Since some fixed label has positive success probability, necessarily $p>0$.
Here $Q_{\mathrm{addr}}$ is the number of queries made by $\mathcal B$, whereas $R_{\mathrm{addr}}$ bounds the number of queries allowed in~\eqref{tot:eq:address-hardness}.

Table~\ref{tot:tab:notation} summarizes the main parameters of the totalization.
We introduce the components of each column and the local notation for the algorithms as needed.
\begin{table}[!htbp]
\centering
\caption{Main notation for the construction of total functions.}
\label{tot:tab:notation}
\renewcommand{\arraystretch}{1.15}
\begin{tabularx}{\textwidth}{@{}p{0.28\textwidth}X@{}}
\toprule
Symbol & Meaning \\
\midrule
$P$ & Function mapping addresses to labels. \\
$a$ & Number of bits in one address input. \\
$K$ & Number of labels and of pointer cells in each column. \\
$m$ & Number of input columns. \\
$b$ & Number of data bits in each column. \\
$Q_{\mathrm{addr}}$ & Number of queries made by the quantum algorithm for $P$. \\
$R_{\mathrm{addr}}$ & Maximum number of queries allowed in~\eqref{tot:eq:address-hardness}. \\
$p$ & Bound on the success probability in~\eqref{tot:eq:address-hardness}. \\
$F_{\mathrm{tot}}$ & Constructed total Boolean function. \\
\bottomrule
\end{tabularx}
\end{table}
\FloatBarrier

\subsection{The Total Function}

We define ${F_{\mathrm{tot}}}$ and show that at most one column can be a winner.
Both the quantum algorithm and the randomized lower bound below refer to this definition.

\begin{definition}[Totalization]\label{tot:def:function}
Let $m\ge2$ be an integer, and let $b\ge2$ be a power of two.
The input $X$ consists of $m$ columns.
Column $j$ contains a data string $y_j\in\{0,1\}^b$, an original address $x_j\in\{0,1\}^a$, and $K$ pointer cells.
Cell $z\in[K]$ stores a pointer $\pi_j(z,i)\in[b]$ for every $i\in[m]\setminus\{j\}$, each represented by $\log b$ bits.
If $x_j\in\dom(P)$, we call cell $P(x_j)$ the \emph{true cell} of column $j$.
In this case, column $j$ is a \emph{winner} if
\begin{enumerate}
 \item Every data bit in column $j$ is one.
 \item The true cell of column $j$ points to a data bit equal to zero in every other column.
\end{enumerate}
A column whose address lies outside $\dom(P)$ is not a winner.
Define ${F_{\mathrm{tot}}}(X)=1$ if and only if $X$ has a winner.
\end{definition}

The requirement that $b$ be a power of two ensures that every pointer encoding represents a position in $[b]$.
The definition therefore assigns a function value to every input bit string.
Its input length is
\begin{equation}\label{tot:eq:input-size}
 n=m\bigl[b+a+K(m-1)\log b\bigr].
\end{equation}
Figure~\ref{fig:winner-construction} depicts a winning column and its true pointer cell.

\Needspace{6\baselineskip}
\begin{lemma}[Uniqueness of the winner]\label{tot:lem:unique}
Every input $X$ has at most one winner.
\end{lemma}
\begin{proof}
Suppose that distinct columns $i,j$ are both winners.
The winning condition for column $i$ gives
\[
 y_{j,\pi_i(P(x_i),j)}=0,
\]
whereas the winning condition for column $j$ requires $y_j=1^b$, a contradiction.
\end{proof}

\subsection{Exact Quantum Query Complexity}

We first select and verify a candidate column coherently, showing that the acceptance probability is determined by the winner's original address.
We then use the address that has been read to rescale this probability and obtain an exact algorithm.

For each valid address $x$, write the success probability of the fixed algorithm $\mathcal B$ as
\begin{equation}\label{tot:eq:seed-success}
 \sigma(x)=\Pr[\mathcal B(x)=P(x)]\ge\frac23.
\end{equation}
The value $\sigma(x)$ is determined by $P$ and $\mathcal B$; it is distinct from the parameter $p$ in the randomized recovery bound.

\begin{lemma}[Candidate selection and verification]\label{tot:lem:verification}
There is a fixed unitary circuit $V$ using $O(mQ_{\mathrm{addr}}+m\log b+a+b)$ queries, with an acceptance flag $A$, such that the following holds.
If ${F_{\mathrm{tot}}}(X)=0$, then $A$ is always zero.
If ${F_{\mathrm{tot}}}(X)=1$ and $w$ is its unique winner, then
\[
 \Pr[A=1]=\sigma(x_w).
\]
On every branch in the computational basis with $A=1$, the register storing the original address contains $x_w$.
\end{lemma}
\begin{proof}
The circuit has three steps, retaining all query answers and intermediate records in its work registers.
\begin{enumerate}
 \item \emph{Decode the labels.}
 Run $\mathcal B$ coherently on every column, using separate work registers, and retain the label registers $Z_1,\ldots,Z_m$.
 Record intermediate measurements coherently.
 This step uses $mQ_{\mathrm{addr}}$ queries.
 \item \emph{Select a candidate.}
 Perform $m-1$ pairwise comparisons according to a fixed elimination schedule.
 On a branch whose labels are $z=(z_1,\ldots,z_m)$, a comparison of columns $i,j$ reads $\pi_i(z_i,j)$ and $\pi_j(z_j,i)$ and queries the data bits to which they point.
 If exactly one of the two columns points to a zero in the other column, retain that column; otherwise, break the tie by a fixed rule.
 Each comparison uses $O(\log b)$ queries, leaving one candidate column $c$.
 \item \emph{Verify the candidate.}
 Read $x_c,y_c$ in full.
 If $x_c$ is valid, compute $P(x_c)$ and read all pointers in its true cell together with the data bits to which they point.
 Set $A=1$ if and only if $c$ satisfies the winning conditions in Definition~\ref{tot:def:function} and $z_c=P(x_c)$.
 If $x_c$ is invalid, use a fixed placeholder label to perform the same number of reads and set $A=0$.
 Validity and $P(x_c)$ are computed from a fixed lookup table.
 This step uses $O(a+b+m\log b)$ queries.
\end{enumerate}
All three steps can be implemented reversibly, retaining the comparison results and leaving the label registers unchanged.
The saved records may coherently control the query positions, but every branch makes the same number of bit queries.
The total cost is
\[
 mQ_{\mathrm{addr}}+O(m\log b)+O(a+b+m\log b)
 =O(mQ_{\mathrm{addr}}+m\log b+a+b).
\]
If ${F_{\mathrm{tot}}}(X)=0$, no candidate satisfies the winning conditions, so verification always rejects.

Now fix an input $X$ with ${F_{\mathrm{tot}}}(X)=1$, and let $w$ be its unique winner.
Given decoded labels $z$, the last two steps are deterministic; denote their final acceptance bit by $a_X(z)$.
Suppose that $z_w=P(x_w)$.
For every opponent $j\ne w$, the two data bits read in a comparison with $w$ satisfy
\[
 y_{j,\pi_w(z_w,j)}=0,
 \qquad y_{w,\pi_j(z_j,w)}=1.
\]
Thus $w$ wins every comparison in which it participates, regardless of $z_j$.
Induction along the fixed schedule shows that $w$ survives every elimination and becomes the final candidate.
It then passes the checks of the winning conditions and the decoded label.
Conversely, $a_X(z)=1$ implies that the candidate is the unique winner $w$ and that its label is correct.
Hence
\begin{equation}\label{tot:eq:accept-equivalence}
 a_X(z)=1\quad\Longleftrightarrow\quad z_w=P(x_w).
\end{equation}

Write the quantum state after decoding as
\[
 \sum_{z\in[K]^m}|z\rangle_Z|\psi_{X,z}\rangle,
 \qquad \sum_z\|\psi_{X,z}\|^2=1,
\]
where the states $|\psi_{X,z}\rangle$ need not be normalized.
A measurement of $Z$ at this point would satisfy $\Pr[Z=z]=\|\psi_{X,z}\|^2$; we use this measurement only for the analysis.
The subsequent computation retains $Z$ and writes $a_X(z)$ into the acceptance flag on the entire $z$ branch.
The branches remain orthogonal, so the acceptance probability is
$\sum_z a_X(z)\|\psi_{X,z}\|^2$.
Substituting~\eqref{tot:eq:accept-equivalence} gives
\begin{equation}\label{tot:eq:accept-probability}
 \begin{aligned}
 \Pr[A=1]
 &=\sum_{z:\,z_w=P(x_w)}\Pr[Z=z]\\
 &=\Pr[Z_w=P(x_w)]=\sigma(x_w).
 \end{aligned}
\end{equation}
Every accepting branch has candidate $w$, so its register storing the original address contains $x_w$.
\end{proof}

\begin{lemma}[Exact quantum upper bound]\label{tot:lem:exact-upper}
The function ${F_{\mathrm{tot}}}$ in Definition~\ref{tot:def:function} has exact quantum query complexity
$O(mQ_{\mathrm{addr}}+m\log b+a+b)$.
\end{lemma}
\begin{proof}
The verification circuit has an acceptance probability determined by the original address stored on every accepting branch.
We use that address to rescale the probability to $1/4$ on positive inputs and then apply exact amplitude amplification.

Append a work qubit $B$, initialized to $|0\rangle$, to the circuit of Lemma~\ref{tot:lem:verification}.
Following~\cite[Section~7]{AIK26}, we rescale the acceptance probability using a rotation $R_x$ on one qubit, fixed in advance for each $x\in\{0,1\}^a$.
For valid $x$, we have $1/(4\sigma(x))\in[1/4,3/8]$, so we may take
\[
 R_x=\begin{pmatrix}
 \sqrt{1-\frac1{4\sigma(x)}}&-\sqrt{\frac1{4\sigma(x)}}\\[0.4em]
 \sqrt{\frac1{4\sigma(x)}}&\sqrt{1-\frac1{4\sigma(x)}}
 \end{pmatrix}.
\]
For $x\notin\dom(P)$, set $R_x=I$.
Control these rotations on the acceptance flag and the register storing the original address, obtaining the fixed unitary
\begin{equation}\label{tot:eq:fixed-calibration}
 U=|0\rangle\langle0|_A\otimes I
   +|1\rangle\langle1|_A\otimes
     \sum_{x\in\{0,1\}^a}|x\rangle\langle x|\otimes R_x.
\end{equation}
This unitary acts as the identity on all other registers.
The resulting circuit is $W=U(V\otimes I_B)$; when $A=0$, the qubit $B$ remains zero.
All values $\sigma(x)$ and rotations $R_x$ are determined by the fixed $P,\mathcal B$ and are constants in the algorithm description.
Implementing~\eqref{tot:eq:fixed-calibration} requires no input queries.

Define the accepting subspace by the fixed projector
\[
 \Pi=|1\rangle\langle1|_A\otimes|1\rangle\langle1|_B\otimes I_{\mathrm{rest}}.
\]
If ${F_{\mathrm{tot}}}(X)=0$, every branch has $A=0$, so the probability of this subspace is zero.
If ${F_{\mathrm{tot}}}(X)=1$, Lemma~\ref{tot:lem:verification} guarantees that every branch with $A=1$ uses the same rotation $R_{x_w}$.
The probability of the accepting subspace is therefore
\begin{equation}\label{tot:eq:calibrated-acceptance}
 \Pr[A=B=1]
 =\sigma(x_w)\frac1{4\sigma(x_w)}=\frac14.
\end{equation}
Combining the two cases, for every input $X$ we have
\[
 \|\Pi W_X|0\rangle\|^2=\frac{{F_{\mathrm{tot}}}(X)}4.
\]

By Lemma~\ref{prelim:lem:exact-amplification}, three calls to $W$ or $W^\dagger$ give an exact algorithm.
The number of queries increases by only a constant factor.
\end{proof}

\subsection{Randomized Query Lower Bound}

We turn one column of a negative input into a winner, obtaining a positive input that differs only in that hidden column.
To distinguish the inputs, an algorithm must find either a modified data bit or the true pointer cell.
We separately bound the probabilities of finding the modified data bit, making more than $R_{\mathrm{addr}}$ queries to the hidden address, and finding its true cell with at most $R_{\mathrm{addr}}$ such queries.

\begin{lemma}[Coupling by planting a winner, adapted from {\cite[Proposition~4.2]{OAI26}}]\label{tot:lem:coupling}
There are coupled inputs $X^-,X^+$ supported on ${F_{\mathrm{tot}}}^{-1}(0)$ and ${F_{\mathrm{tot}}}^{-1}(1)$, respectively, such that every randomized algorithm $A$ making at most $T$ queries satisfies
\begin{equation}\label{tot:eq:coupling}
 \left|\Pr[A(X^+)=1]-\Pr[A(X^-)=1]\right|
 \le \frac{4T}{mb}+\frac{T}{mR_{\mathrm{addr}}}+Tp.
\end{equation}
\end{lemma}
\begin{proof}
We construct a coupling that changes only one hidden data bit and one pointer cell.
We bound the probability of detecting either change by considering data queries and queries to the hidden address separately.

For each column $j$, sample $J_j\in[b]$ independently and uniformly, and independently sample $x_j\sim\mu$.
Set the data and pointers of the negative input $X^-$ to
\[
 y_{j,t}^- =1-\mathbf 1_{\{t=J_j\}},
 \qquad \pi_j^-(z,i)=1\quad(i\ne j,\ z\in[K]).
\]
Every column contains a zero, so ${F_{\mathrm{tot}}}(X^-)=0$.
Independently choose a uniform $I\in[m]$, set $y_{I,J_I}^+=1$, and set
\[
 \pi_I^+(P(x_I),j)=J_j\qquad(j\ne I).
\]
Leave all other input bits unchanged to obtain $X^+$.
Since $x_I\in\dom(P)$, column $I$ becomes a winner, and hence ${F_{\mathrm{tot}}}(X^+)=1$.
The two inputs can differ only at $(I,J_I)$ and inside the true pointer cell of column $I$.

Run $A$ on the two inputs with the same random coins, and denote its execution records, including the outputs, by $\tau^-,\tau^+$.
Let $E_{\rm hit}$ be the event that the execution on $X^-$ queries $(I,J_I)$ or some bit in that true pointer cell.
Before the first such query, the two inputs give identical answers.
Induction on the query steps shows that if $E_{\rm hit}$ does not occur, the records are identical.
Thus the coupling inequality gives
\[
 \begin{aligned}
 \left|\Pr[A(X^+)=1]-\Pr[A(X^-)=1]\right|
 &\le\TV(\mathcal L(\tau^+),\mathcal L(\tau^-))\\
 &\le\Pr[\tau^+\ne\tau^-]\le\Pr[E_{\rm hit}],
 \end{aligned}
\]
where $\mathcal L$ denotes the distribution of a random variable.
All query counts and events below are defined along the execution on $X^-$.

\begin{enumerate}
 \item \emph{Hitting the deleted zero.}
 Let $q_j^{\rm d}$ be the number of distinct queried data positions in column $j$, let $N$ be the number of columns in which a zero is found, and put
 \[
 \mathcal H_{\mathrm{data}}=\{j:q_j^{\rm d}>b/2\}.
 \]
 Since $\sum_jq_j^{\rm d}\le T$, every execution satisfies $|\mathcal H_{\mathrm{data}}|\le2T/b$.
 Let $Z_s$ indicate that step $s$ first discovers the zero in some column and that at most $b/2$ distinct data positions in that column were queried before this step.
 Every discovered zero either belongs to a column in $\mathcal H_{\mathrm{data}}$ or is counted by some $Z_s$, so
 \[
 N\le|\mathcal H_{\mathrm{data}}|+\sum_{s=1}^T Z_s.
 \]
 Condition on the record preceding step $s$.
 If this step queries a new data bit in a column $j$ whose zero has not yet been found, then $J_j$ remains uniform among the unqueried positions of that column.
 Indeed, the address instances are independent of all $J_j$, and the fixed pointer table in $X^-$ provides no information about the zero positions.
 If $r\le b/2$ distinct positions in that column have already been queried, the conditional probability of hitting its zero is $1/(b-r)\le2/b$; in all other cases, $Z_s=0$.
 Consequently,
 \[
 \E N\le\frac{2T}{b}+\sum_{s=1}^T\E Z_s
 \le\frac{4T}{b}.
 \]
 Let $E_{\rm d}$ be the event that $(I,J_I)$ is queried.
 Since $I$ is independent of the entire execution on $X^-$,
 \[
 \Pr[E_{\rm d}]=\frac{\E N}{m}\le\frac{4T}{mb}.
 \]

 \item \emph{Making more than $R_{\mathrm{addr}}$ queries to the target address.}
 Let $q_i^{\rm a}$ count the queries to address bits in column $i$ over the full execution, and put
 \[
 \mathcal H_{\mathrm{addr}}=\{i:q_i^{\rm a}>R_{\mathrm{addr}}\}.
 \]
 Since $\sum_iq_i^{\rm a}\le T$, we have
 $|\mathcal H_{\mathrm{addr}}|\le T/(R_{\mathrm{addr}}+1)\le T/R_{\mathrm{addr}}$.
 Independence and uniformity of $I$ give
 \[
 \Pr[I\in\mathcal H_{\mathrm{addr}}]
 =\frac{\E|\mathcal H_{\mathrm{addr}}|}{m}\le\frac{T}{mR_{\mathrm{addr}}}.
 \]

 \item \emph{Accessing the true cell after at most $R_{\mathrm{addr}}$ address queries.}
 Let $E_{i,s}$ be the event that step $s$ queries a bit in cell $P(x_i)$ of column $i$ and that at most $R_{\mathrm{addr}}$ queries to $x_i$ precede this step.
 Fix $i,s$, the algorithm's random coins, all $J_j$, and all address instances other than $x_i$.
 These quantities are independent of $x_i$, so the remaining input $x_i$ still has distribution $\mu$.

 Construct a decision tree $B_{i,s}$ on $x_i$ as follows.
 Simulate $A$ on $X^-$, answering every query outside $x_i$ from the fixed information.
 Stop immediately before a possible $(R_{\mathrm{addr}}+1)$st query to $x_i$.
 If step $s$ is reached and queries cell $z$ in column $i$, output $z$; otherwise, output a fixed label in $[K]$.
 This decision tree makes at most $R_{\mathrm{addr}}$ queries on every $x_i$.

 Whenever $E_{i,s}$ occurs, the tree reaches step $s$ and outputs $P(x_i)$.
 Writing $\omega$ for the tuple of fixed quantities,~\eqref{tot:eq:address-hardness} gives
 \[
 \Pr[E_{i,s}\mid\omega]
 \le\Pr[B_{i,s}(x_i)=P(x_i)\mid\omega]
 \le p.
 \]
 Averaging over $\omega$ yields $\Pr[E_{i,s}]\le p$.
 Independence of $I$ and the union bound now give
 \[
 \begin{aligned}
 \Pr\!\left[\bigcup_{s=1}^T E_{I,s}\right]
 &=\frac1m\sum_{i=1}^m
     \Pr\!\left[\bigcup_{s=1}^T E_{i,s}\right]\\
 &\le\frac1m\sum_{i=1}^m\sum_{s=1}^T\Pr[E_{i,s}]\le Tp.
 \end{aligned}
 \]
 The recovery bound is applied before conditioning on any event involving $\mathcal H_{\mathrm{addr}}$.
\end{enumerate}

If $E_{\rm hit}$ occurs through a pointer query and $I\notin\mathcal H_{\mathrm{addr}}$, that query is counted by some $E_{I,s}$.
Therefore,
\[
 E_{\rm hit}\subseteq E_{\rm d}\cup\{I\in\mathcal H_{\mathrm{addr}}\}
                         \cup\bigcup_{s=1}^T E_{I,s}.
\]
Substituting the three probability estimates into the bound for the coupled execution records proves~\eqref{tot:eq:coupling}.
\end{proof}

\subsection{The Totalization Theorem}

Combining the quantum algorithm with the randomized coupling bound, and using an AND restriction for the exact quantum lower bound, gives the following totalization theorem.
The number of labels $K$ affects the input length but does not appear explicitly in the query bounds.

\begin{theorem}[Totalization with multiple address labels]\label{tot:prop:totalization}
There are absolute constants $c,C>0$ such that the total function in Definition~\ref{tot:def:function} satisfies
\begin{equation}\label{tot:eq:exact-total}
 b\le\QE({F_{\mathrm{tot}}})\le C(mQ_{\mathrm{addr}}+m\log b+a+b)
\end{equation}
and
\begin{equation}\label{tot:eq:random-total}
 \R({F_{\mathrm{tot}}})\ge c\min\{mb,mR_{\mathrm{addr}},p^{-1}\}.
\end{equation}
The constants are independent of $K$.
The quantum upper bound holds in the standard query model with arbitrary exact unitaries that do not depend on the input.
\end{theorem}
\begin{proof}
We combine the verified exact algorithm with an AND restriction to obtain the quantum bounds.
The randomized bound follows by applying the coupling lemma to an algorithm with error at most $1/3$.

Lemma~\ref{tot:lem:exact-upper} gives the quantum upper bound.
For the lower bound, fix a valid address $x^\star$, set every column's address to $x^\star$, and make every pointer point to the first data bit of its target column.
Fix the data in columns $2,\ldots,m$ to $0^b$, leaving only the data string $u\in\{0,1\}^b$ of column $1$ free.
No other column is a winner, and column $1$ is a winner exactly when $u=1^b$.
This restriction is therefore $\AND_b$.
Equation~\eqref{prelim:eq:and-restriction} implies $\QE(F_{\mathrm{tot}})\ge b$, proving~\eqref{tot:eq:exact-total}.

Suppose that a randomized algorithm $A$ makes at most $T$ queries and computes ${F_{\mathrm{tot}}}$ with probability at least $2/3$ on every input.
Since $X^+$ and $X^-$ are supported on ${F_{\mathrm{tot}}}^{-1}(1)$ and ${F_{\mathrm{tot}}}^{-1}(0)$, respectively, averaging the pointwise correctness guarantee gives
\[
 \Pr[A(X^+)=1]\ge\frac23,
 \qquad \Pr[A(X^-)=1]\le\frac13.
\]
Lemma~\ref{tot:lem:coupling} therefore yields
\[
 \frac13\le\frac{4T}{mb}+\frac{T}{mR_{\mathrm{addr}}}+Tp.
\]
All three denominators are at least $\min\{mb,mR_{\mathrm{addr}},p^{-1}\}$, so
\[
 T\ge\frac{1}{3\bigl(4/(mb)+1/(mR_{\mathrm{addr}})+p\bigr)}
 \ge\frac1{18}\min\{mb,mR_{\mathrm{addr}},p^{-1}\}.
\]
Taking the minimum number of queries over all such algorithms proves~\eqref{tot:eq:random-total}.
\end{proof}

\subsection{Forrelation Addresses}\label{tot:app:simple}\label{tot:sec:main}

To apply Theorem~\ref{tot:prop:totalization} to $k$-Forrelation, we use the answers to several independent instances as the label function $P$.
We bound the quantum query complexity of recovering the full label and the success probability of randomized algorithms making at most $R_{\mathrm{addr}}$ queries.
These determine the parameters in Theorem~\ref{tot:prop:totalization}.

Let $d_0=2^h$ for an integer $h\ge1$, and let $k\ge2$ be an integer.
Let $H_{d_0}$ be the normalized Hadamard matrix and $u=d_0^{-1/2}\sum_i|i\rangle$.
The input consists of $k$ sign vectors $z_1,\ldots,z_k\in\{\pm1\}^{d_0}$.
Writing $D_j=\operatorname{diag}(z_j)$, define
\[
 \Phi_{k,d_0}(z)=\langle u|D_1H_{d_0}D_2\cdots H_{d_0}D_k|u\rangle,
 \qquad \delta=2^{-5k}.
\]
Here $\Phi_{k,d_0}$ is the $k$-Forrelation quantity.
Define the corresponding $k$-Forrelation promise function $\operatorname{Forr}_{k,d_0}$ to output $1$ when $\Phi_{k,d_0}\ge\delta$ and $0$ when $|\Phi_{k,d_0}|\le\delta/2$.
All other inputs lie outside its promise domain.

\begin{lemma}[Bansal--Sinha {\cite[Corollary~1.4]{BS21}}]
\label{thm:forrelation-lower}
Let $k\ge2$ be an integer, and let $d_0\ge2^{15k+2}$ be a power of two.
The partial function $\operatorname{Forr}_{k,d_0}$ has a fixed quantum algorithm with error probability at most $1/3$ using $O(k2^{10k})$ queries, and satisfies
\[
 \R(\operatorname{Forr}_{k,d_0})\ge c_* k^{-29}
 \frac{d_0^{1-1/k}}{\log(kd_0)},
\]
where $c_*>0$ is an absolute constant independent of $k,d_0$.
\end{lemma}

We use a weaker form of the cited result.
The condition $d_0\ge2^{15k+2}$ gives $4/(d_0\delta^2)\le\delta$, as required in~\cite[Section~5.4]{BS21}; the constants in its Fourier weight estimate are uniform in $k$.
All parameters below are chosen in this range.
Fix a sufficiently small absolute constant $\gamma>0$, and set
\[
 R_{\mathrm{addr}}=\left\lfloor\gamma k^{-29}\frac{d_0^{1-1/k}}{\log(kd_0)}\right\rfloor.
\]
The choices below ensure $R_{\mathrm{addr}}\ge1$.
By Lemma~\ref{prelim:lem:yao}, there is a distribution $\nu$ on valid inputs under which every randomized algorithm making at most $R_{\mathrm{addr}}$ queries succeeds with probability at most $2/3$.

Let $r=10h$, and let $P$ output the label formed by the answers to $r$ instances of $\operatorname{Forr}_{k,d_0}$.
Its input length and number of labels are
\[
 a=rkd_0,\qquad K=2^r=d_0^{10}.
\]
For the randomized lower bound, the instances are independent samples from $\nu$.
Fix
\[
 \mu=\nu^{\otimes r},\qquad p=(2/3)^r.
\]

\begin{corollary}[Recovery of address labels]\label{tot:lem:forrelation-address}
There is a quantum algorithm that makes
\[
 Q_{\mathrm{addr}}=\Theta(kr2^{10k}\log r)
\]
queries and recovers the full label with probability at least $2/3$ on every valid input.
Under $\mu$, every randomized algorithm making at most $R_{\mathrm{addr}}$ queries in total recovers the full label with probability at most $p\le d_0^{-5}$.
\end{corollary}
\begin{proof}
For the quantum upper bound, amplify the individual answers and recover all labels.
For the randomized lower bound, apply the direct product lemma to the independent address instances.

By Lemma~\ref{thm:forrelation-lower}, each instance has a quantum algorithm with error probability at most $1/3$ using $O(k2^{10k})$ queries.
Repeat this algorithm independently $O(\log r)$ times and take the majority answer.
The Chernoff bound~\cite{Chernoff52} reduces the error probability to $1/(3r)$.
The number of queries is fixed, including outside the promise domain.
Retaining the intermediate results and computing the majority reversibly gives a unitary circuit with the same output distribution.
Process the $r$ instances in turn and, if necessary, pad with queries that do not affect the output, so that the total number of queries is
$Q_{\mathrm{addr}}=\Theta(kr2^{10k}\log r)$.
By the union bound, the probability that the full label is incorrect is at most $r/(3r)=1/3$.

Fix any randomized algorithm $A$ making at most $R_{\mathrm{addr}}$ queries in total, and let
$x=(x^{(1)},\ldots,x^{(r)})\sim\mu$.
It also makes at most $R_{\mathrm{addr}}$ queries to each instance, so Lemma~\ref{lem:local-budget-product} gives
\[
 \Pr_{x\sim\mu}[A(x)=P(x)]\le p\le d_0^{-5}.
\]
The probability also includes the internal randomness of $A$.
The last inequality follows from
\[
 (2/3)^r=2^{-10h\log(3/2)}
 =d_0^{-10\log(3/2)}\le d_0^{-5},
\]
since $10\log(3/2)>5$.
\end{proof}

\subsection{A Nearly Quadratic Separation}

We substitute the bounds for address recovery into the totalization theorem and use $k$ to balance the quantum recovery cost against the loss in the randomized lower bound.
This gives the following nearly quadratic separation.

\begin{theorem}\label{tot:thm:query}
There are a constant $C>0$ and a family of total Boolean functions $F_{\mathrm{tot},h}$ indexed by all sufficiently large integers $h$.
Let $t_h:=Q_E(F_{\mathrm{tot},h})$.
Their input lengths satisfy $n_h=t_h^{O(1)}$, and
\[
 R(F_{\mathrm{tot},h})\ge t_h^2\,2^{-C\sqrt{\log t_h}}=t_h^{2-o(1)}.
\]
Each $F_{\mathrm{tot},h}$ has an $\AND_{b_h}$ restriction with $b_h=\Theta(t_h)$.
\end{theorem}
\begin{proof}
We balance the Forrelation order against the cost of address recovery, and choose the data length so that the AND lower bound matches the exact algorithm.
We then express the randomized lower bound and input length in terms of the actual exact quantum complexity.

Set $d_0=2^h$, $k=\lceil\sqrt h\rceil$, and $m=d_0$, and use the address construction from Corollary~\ref{tot:lem:forrelation-address}.
For sufficiently large $h$, we have $k\ge2$ and $h\ge15k+2$, as required by Lemma~\ref{thm:forrelation-lower}.
Let $b_h$ be the smallest power of two at least $d_0Q_{\mathrm{addr}}$, and apply the totalization theorem with $b=b_h$.
We estimate the exact quantum complexity, the randomized complexity, and the input length in turn.

Since $r=10h$, the number of queries for address recovery satisfies
\[
 \log Q_{\mathrm{addr}}=10k+O(\log h),\qquad
 d_0Q_{\mathrm{addr}}\le b_h<2d_0Q_{\mathrm{addr}}.
\]
Moreover, $a=rkd_0=O(d_0Q_{\mathrm{addr}})$, and
$\log b_h=O(h)$ and $Q_{\mathrm{addr}}\ge h$ imply $m\log b_h=O(d_0Q_{\mathrm{addr}})$.
The quantum upper and lower bounds in Theorem~\ref{tot:prop:totalization} give
\[
 t_h=\Theta(d_0Q_{\mathrm{addr}}),\qquad
 \log t_h=h+10k+O(\log h),\qquad b_h=\Theta(t_h).
\]
The restriction used in its proof also shows that $F_{\mathrm{tot},h}$ has an $\AND_{b_h}$ restriction.

Since $k=\Theta(\sqrt h)$, the quantity rounded down in the definition of $R_{\mathrm{addr}}$ is $2^{h-O(\sqrt h)}$ and tends to infinity.
Thus rounding loses at most a constant factor, and
\[
 \log R_{\mathrm{addr}}
 =h-\frac hk-29\log k-\log\log(kd_0)+O(1)
 =h-\frac hk+O(\log h).
\]
Its definition also gives $R_{\mathrm{addr}}\le d_0\le b_h$, while
$mR_{\mathrm{addr}}p\le d_0^2d_0^{-5}<1$.
The minimum in the totalization theorem is therefore $mR_{\mathrm{addr}}$, giving
\[
 \log R(F_{\mathrm{tot},h})\ge2h-\frac hk-O(\log h).
\]
Combining this with $2\log t_h=2h+20k+O(\log h)$ yields
\[
 \log\frac{t_h^2}{R(F_{\mathrm{tot},h})}
 \le20k+\frac hk+O(\log h)=O(\sqrt h).
\]
Since $h\le\log t_h$, this proves the desired nearly quadratic separation.

Finally, $\log Q_{\mathrm{addr}}=O(\sqrt h)$ implies $Q_{\mathrm{addr}}\le d_0$ for sufficiently large $h$.
Hence $a,b_h=O(d_0^2)$.
Substituting $m=d_0$ and $K=d_0^{10}$ into~\eqref{tot:eq:input-size}, we obtain
\[
 n_h=m\bigl[b_h+a+K(m-1)\log b_h\bigr]
 =O(d_0^{12}\log d_0)=t_h^{O(1)},
\]
where the final step uses $t_h\ge d_0$.
\end{proof}

%% file: sections/constant.tex
\section{Partial Functions}
\label{cp-section}

We construct a partial function $F_{\mathrm{par}}$ from $k$-Forrelation~\cite{AA18,BS21}, separating constant exact quantum query complexity from nearly linear randomized query complexity.
The construction has three steps: encode an address in the answers to hard instances, store the exact success probability of the quantum address algorithm at that address, and read this probability to rescale the accepting amplitude.
A randomized algorithm cannot distinguish the two types of input without finding the true cell.
All address instances and the additional table count toward the total input length $n$.

The correctness of the additional probability table is part of the promise.
For each fixed $\varepsilon\in(0,1)$, we choose constants $k,d$ that keep the quantum query complexity constant, and then let the input length grow.
Table~\ref{cp-notation} lists the main parameters and intermediate objects; other notation is introduced where it is used.
\begin{table}[!htbp]
\centering
\caption{Main notation for the construction of the partial function.}
\label{cp-notation}
\renewcommand{\arraystretch}{1.15}
\begin{tabularx}{\textwidth}{@{}p{0.28\textwidth}X@{}}
\toprule
Symbol & Meaning \\
\midrule
$k$ & Order of $k$-Forrelation. \\
$d$ & Number of address blocks. \\
$b_{\mathrm{enc}}$ & Number of bits in each address or probability block. \\
$M$ & Number of addresses and length of each Forrelation table. \\
$J$ & Number of masked Forrelation instances. \\
$\operatorname{Forr}_{k,M}$ & The $k$-Forrelation partial Boolean function. \\
$g$ & The Forrelation function with an additional mask bit. \\
$\ell(X)$ & True address encoded by the base input. \\
$p_D(X)$ & Probability that the circuit recovers the address. \\
$s$ & Number of blocks encoding the success probability in each table cell. \\
$F_{\mathrm{par}}$ & The constructed partial Boolean function. \\
\bottomrule
\end{tabularx}
\end{table}

\subsection{Forrelation Algorithm}
\label{cp-seed}

The address construction requires a quantum algorithm for $k$-Forrelation.
We record both its query count and its Hadamard gate count: the former controls the query cost, while the latter determines the exact denominator of its success probability in Section~\ref{cp-address}.
Fix integers $k\ge2,d\ge1$ and a positive integer ${b_{\mathrm{enc}}}$, and write $M=2^{d{b_{\mathrm{enc}}}}$.
Let $H_M$ be the normalized Walsh--Hadamard matrix.
For $k$ sign tables $z_1,\ldots,z_k\in\{\pm1\}^{M}$, write $Z_i=\operatorname{diag}(z_i)$ and define
\[
 \Phi_{k,M}(z)=\frac1M z_1^{\mathsf T}H_M Z_2 H_M\cdots Z_{k-1}H_M z_k.
\]
When $k=2$, there are no intermediate diagonal matrices.
Let $\Delta=2^{-5k}$.
The $k$-Forrelation problem used in this section is the following partial function on $kM$ bits:
\[
 \operatorname{Forr}_{k,M}(z)=\begin{cases}
 1,&\Phi_{k,M}(z)\ge\Delta,\\
 0,&|\Phi_{k,M}(z)|\le\Delta/2.
 \end{cases}
\]
All other inputs lie outside the promise domain.
For fixed $k$, Lemma~\ref{thm:forrelation-lower} with $d_0=M$ gives, for all sufficiently large $M$,
\begin{equation}\label{cp-seed-lower}
 R(\operatorname{Forr}_{k,M})=\Omega_k\!\left(\frac{M^{1-1/k}}{\log M}\right).
\end{equation}
We first state the basic circuit of Aaronson and Ambainis, and then amplify it to the error required by our promise, keeping track of the gate set and gate count needed to represent its success probability.

\begin{lemma}[Aaronson--Ambainis~{\cite[Section~3.2, Figure~1]{AA18}}]\label{cp-aa18-circuit}
There is a fixed unitary circuit consisting of $k$ phase queries and $k+1$ layers of $H_M$ whose amplitude on the output $|0\rangle$ is $\Phi_{k,M}(z)$ on every input $z$.
\end{lemma}

\begin{corollary}[Amplifying the Forrelation algorithm]\label{cp-seed-algorithm}
The partial function $\operatorname{Forr}_{k,M}$ has a fixed unitary query algorithm with error at most $\eta=1/(100d)$.
Its query count ${q_{\mathrm F}}$ and number ${H_{\mathrm F}}$ of Hadamard gates acting on one qubit satisfy
\[
 {q_{\mathrm F}}=O(k2^{10k}\log(d+1)),\qquad {H_{\mathrm F}}=O({q_{\mathrm F}}\log M).
\]
All other gates are signed permutations of the computational basis.
\end{corollary}
\begin{proof}
We repeat the basic circuit to reduce its error and then implement the decision coherently, keeping track of both queries and Hadamard gates.

The circuit in Lemma~\ref{cp-aa18-circuit} outputs all zeros with probability $|\Phi_{k,M}(z)|^2$.
Under our promise, this probability is at most $\Delta^2/4$ when the function value is zero and at least $\Delta^2$ when it is one.
Run the circuit
\[
 r_0=\lceil C\Delta^{-2}\ln(100d)\rceil
\]
times on independent registers, and decide whether the frequency with which the output is zero is at least $\Delta^2/2$.
The multiplicative Chernoff bound~\cite{Chernoff52} gives error $\exp(-\Omega(r_0\Delta^2))$.
For a sufficiently large absolute constant $C$, this is at most $\eta=1/(100d)$.
Retaining the output of every trial and reversibly computing the threshold decision gives a fixed unitary circuit with the same output probabilities.

Each trial of the basic circuit uses $k+1$ copies of $H_M=H^{\otimes\log M}$.
The standard bit oracle implements the phase queries through a shared $|{-}\rangle$ ancilla; preparing and restoring that ancilla uses two additional Hadamard gates.
All remaining operations, including the tests for output zero and the threshold computation, are signed permutations of the computational basis.
Thus
\[
 \begin{aligned}
 q_{\mathrm F}&=kr_0=O(k2^{10k}\log(d+1)),\\
 H_{\mathrm F}&=r_0\bigl((k+1)\log M+2\bigr)
              =O(q_{\mathrm F}\log M).
 \end{aligned}
\]
Both the number of trials and the circuit are fixed in advance, independently of the input.
\end{proof}

Add a mask bit $z$ to each Forrelation instance and define
\[
 g(x,z)=\operatorname{Forr}_{k,M}(x)\mathbin\oplus z.
\]
The input length of $g$ is $kM+1$.
XORing $z$ into the answer of the preceding algorithm computes $g$ using ${q_{\mathrm F}}+1$ queries, with error still at most $\eta$ and with the same Hadamard count ${H_{\mathrm F}}$.
We will use the mask bit to make the answer uniform under the hard distribution.

\FloatBarrier
\subsection{Address Encoding}
\label{cp-address}

Organizing the answers of the preceding instances into a Hadamard encoding lets a quantum algorithm recover a long address with a query count independent of the address block length ${b_{\mathrm{enc}}}$.
We construct the circuit and represent its success probability exactly, for use in the exact algorithm of the next subsection.

Partition an address $\ell\in\mathbb F_2^{d{b_{\mathrm{enc}}}}$ into $d$ blocks $w_1,\ldots,w_d$ of ${b_{\mathrm{enc}}}$ bits each.
Define the injective linear encoding
\[
 \operatorname{Enc}(\ell)_{j,v}=w_j\cdot v,
 \qquad j\in[d],\quad v\in\mathbb F_2^{b_{\mathrm{enc}}}.
\]
The inner product is over $\mathbb F_2$.
The number of instances, or equivalently the encoding length, is
\[
 J=d2^{b_{\mathrm{enc}}}=dM^{1/d}.
\]
There are $2^{d{b_{\mathrm{enc}}}}=M$ possible addresses.
Thus the number of addresses in this construction equals the length of each Forrelation sign table.
The base input $X=(X_{j,v})$ consists of $J$ valid instances of $g$, each of length $kM+1$, so its total length is $J(kM+1)$.
We promise that the answer vector $(g(X_{j,v}))_{j,v}$ equals $\operatorname{Enc}(\ell)$ for some address $\ell$.
Injectivity makes this address unique; denote it by $\ell(X)$.

Every address has a valid base input.
For each coordinate, choose any valid Forrelation input $x$ and set the mask bit $z$ so that $g(x,z)$ equals the required encoding bit.
Hence the encoding promise is nonempty and excludes no address.

\begin{lemma}\label{cp-address-lemma}
There is a fixed unitary circuit ${\mathcal D}$ that queries only $X$.
Its query count $q_D$ and Hadamard gate count $H_D$ are
\[
 q_D=2d({q_{\mathrm F}}+1),\qquad H_D=d(2{H_{\mathrm F}}+2{b_{\mathrm{enc}}}).
\]
Let ${\mathcal D}(X)$ denote the result of measuring its address output register.
Its success probability satisfies
\begin{equation}\label{cp-probability}
 {p_D}(X):=\Pr[{\mathcal D}(X)=\ell(X)]\ge\frac{24}{25},
 \qquad {p_D}(X)=\frac{u(X)}{2^{H_D}},
\end{equation}
where the integer $u(X)$ has a binary representation of at most $H_D+1$ bits.
\end{lemma}
\begin{proof}
We first analyze the coherent recovery of one address block and combine the blockwise error bounds.
We then count the Hadamard gates to obtain an exact binary representation of the success probability.

Fix a block $j$.
Let $B_{j,v}$ be the algorithm for the instance $X_{j,v}$, including its work registers, and let $Z_{\mathrm{out}}$ apply phases $1$ and $-1$ when the answer bit is zero and one, respectively.
The circuit that recovers block $j$ is
\[
 {\mathcal D}_j=(H^{\otimes {b_{\mathrm{enc}}}}\otimes I)
 \left(\sum_v|v\rangle\langle v|\otimes B_{j,v}^\dagger Z_{\mathrm{out}}B_{j,v}\right)
 (H^{\otimes {b_{\mathrm{enc}}}}\otimes I).
\]
The circuits $B_{j,v}$ have the same operations between queries; only the oracle address has the additional index $(j,v)$.
Consequently, coherent indexing introduces no controlled Hadamard gates.
Different values of $j$ use independent work registers.

Let $e_{j,v}\le\eta$ be the algorithm's error on $X_{j,v}$, whose correct answer is $w_j\cdot v$.
The components of $B_{j,v}|0\rangle$ corresponding to correct and incorrect answers have squared norms $1-e_{j,v}$ and $e_{j,v}$.
They lie in orthogonal answer subspaces, so
\[
 \begin{aligned}
 \langle0|B_{j,v}^\dagger Z_{\mathrm{out}}B_{j,v}|0\rangle
 &=(1-e_{j,v})(-1)^{w_j\cdot v}+e_{j,v}(-1)^{1-w_j\cdot v}\\
 &=(-1)^{w_j\cdot v}(1-2e_{j,v}).
 \end{aligned}
\]
Using $\langle w_j|H^{\otimes {b_{\mathrm{enc}}}}|v\rangle=2^{-{b_{\mathrm{enc}}}/2}(-1)^{w_j\cdot v}$, we obtain
\[
 \langle w_j,0|{\mathcal D}_j|0,0\rangle
 =2^{-{b_{\mathrm{enc}}}}\sum_v(1-2e_{j,v})\ge1-2\eta.
\]
This amplitude corresponds to recovering the correct address with the work register equal to zero.
Dropping the condition on the work register can only increase the probability.
If $\widehat w_j$ denotes the measured index register, then
\[
 \Pr[\widehat w_j\ne w_j]\le1-(1-2\eta)^2\le4\eta,
 \qquad
 \Pr[{\mathcal D}(X)\ne\ell(X)]\le\sum_{j=1}^d4\eta=\frac1{25}.
\]
Each block calls the instance algorithm and its inverse once each, and applies two Hadamard transforms to the index register.
Hence $q_D=2d({q_{\mathrm F}}+1)$ and $H_D=d(2{H_{\mathrm F}}+2{b_{\mathrm{enc}}})$.

For fixed $X$, every gate other than these $H_D$ Hadamard gates is a signed permutation, including the bit oracle.
Starting from a computational basis state, induction on the number $t$ of Hadamard gates applied shows that every amplitude is an integer multiple of $2^{-t/2}$: a Hadamard gate replaces two integer numerators by their sum and difference, while a signed permutation only changes their positions and signs.
The final state therefore has the form
\[
 {\mathcal D}_X|0\rangle
 =2^{-H_D/2}\sum_{\lambda,r}c_{\lambda,r}(X)|\lambda,r\rangle,
 \qquad c_{\lambda,r}(X)\in\mathbb Z,
\]
where $r$ ranges over the computational basis of the work register.
It follows that
\[
 {p_D}(X)=2^{-H_D}\sum_r c_{\ell(X),r}(X)^2,
 \qquad
 u(X):=\sum_r c_{\ell(X),r}(X)^2\in\{0,\ldots,2^{H_D}\}.
\]
The upper bound follows from normalization of the final state, so the numerator requires at most $H_D+1$ bits.
\end{proof}

\subsection{Exact Quantum Algorithm}
\label{cp-exactification}

The preceding subsection gives a circuit ${\mathcal D}$ that recovers the address with success probability ${p_D}(X)=u(X)/2^{H_D}$.
We store an encoding of $u(X)$ at the true address, so that the algorithm can read this information and rescale its accepting probability to exactly $1/4$.
This rescaling uses the weighting method of~\cite[Section~7]{AIK26}; related uses of Hadamard encoding and Bernstein--Vazirani decoding appear in~\cite[Section~1.2]{GS26}.

Identify the address space $\mathbb F_2^{d{b_{\mathrm{enc}}}}$ with $[M]$ through a fixed bijection, and continue to denote the true address by $\ell(X)$.
Let
\[
 s=\left\lceil\frac{H_D+1}{{b_{\mathrm{enc}}}}\right\rceil.
\]
Pad the binary representation of $u(X)$ with zeros to length $s{b_{\mathrm{enc}}}$ and partition it into $s$ blocks of ${b_{\mathrm{enc}}}$ bits.
Encode each block $c\in\mathbb F_2^{b_{\mathrm{enc}}}$ by its Hadamard codeword
$(c\cdot v)_{v\in\mathbb F_2^{b_{\mathrm{enc}}}}$ of length $2^{b_{\mathrm{enc}}}$.
The additional table $Y$ has $M$ cells, each containing one flag bit and $s$ encoding tables.

\begin{definition}[The partial function $F_{\mathrm{par}}$]\label{cp-function}
An input to ${F_{\mathrm{par}}}$ is a pair $(X,Y)$, where $X$ satisfies the encoding promise of Section~\ref{cp-address} and $Y$ satisfies one of the following two promises:
\begin{itemize}
 \item ${F_{\mathrm{par}}}(X,Y)=0$: all of $Y$ is zero;
 \item ${F_{\mathrm{par}}}(X,Y)=1$: cell $\ell(X)$ contains flag $1$ and the encoding of $u(X)$, and every other cell is zero.
\end{itemize}
\end{definition}
The correctness of the probability encoding is part of the promise.
The circuit ${\mathcal D}$ is fixed before $Y$ is defined and queries only $X$, so neither ${p_D}(X)$ nor $u(X)$ depends on $Y$.
The table $Y$ is part of the input, and the total input length of ${F_{\mathrm{par}}}$ is
\begin{equation}\label{cp-length}
 n=J(kM+1)+M(1+s2^{b_{\mathrm{enc}}}).
\end{equation}

\begin{lemma}[Exact computation from the probability encoding]\label{cp-exactification-lemma}
The partial function ${F_{\mathrm{par}}}$ in Definition~\ref{cp-function} satisfies
\begin{equation}\label{cp-q-cost}
 Q_E({F_{\mathrm{par}}})\le3(q_D+1+s).
\end{equation}
\end{lemma}
\begin{proof}
We use the candidate address to read the flag and the encoded success probability, and then apply a fixed controlled rotation.
The promise makes the resulting acceptance probability exactly zero or $1/4$, allowing one round of exact amplitude amplification.

Retain the candidate address register and write the state after running ${\mathcal D}$ as
\[
 {\mathcal D}_X|0\rangle=\sum_{\lambda=1}^M|\lambda\rangle|\psi_\lambda(X)\rangle,
 \qquad
 \sum_\lambda\|\psi_\lambda(X)\|^2=1,
 \qquad
 \|\psi_{\ell(X)}(X)\|^2={p_D}(X).
\]
Let $\chi_\lambda$ be the flag of cell $\lambda$, and let $v_\lambda$ be the integer represented by its $s$ encoded blocks.
One flag query and $s$ applications of the decoding algorithm in Lemma~\ref{prelim:lem:bv} coherently append $\chi_\lambda,v_\lambda$ to each branch.
On a promised input, all encoded blocks are valid Hadamard codewords, including the zero codewords in unused cells.
Every branch makes the same $1+s$ queries, regardless of its flag.

For each integer $v\in\{0,\ldots,2^{s{b_{\mathrm{enc}}}}-1\}$, define
\[
 \omega(v)=
 \begin{cases}
 2^{H_D}/(4v),&1/4\le v/2^{H_D}\le1,\\
 0,&\text{otherwise}.
 \end{cases}
\]
All these values lie in $[0,1]$.
For each $v$, fix in advance a unitary $R_v$ on one qubit satisfying
\[
 R_v|0\rangle=\sqrt{1-\omega(v)}\,|0\rangle
                  +\sqrt{\omega(v)}\,|1\rangle.
\]
Apply $R_v$ to a fresh accepting ancilla if and only if the retained flag equals one.
Equivalently, this controlled rotation acts as $I$ on the subspace with flag zero and as $R_v$ on the subspace with flag one and decoded value $v$.
The entire unitary matrix depends only on the fixed integers $H_D,s,{b_{\mathrm{enc}}}$, not on the input.
It is therefore an allowed fixed exact unitary and requires no queries.
Call the resulting circuit $W$, and let $\Pi$ project onto the subspace in which the accepting ancilla equals one.

The candidate address register is unchanged, so branches for different addresses remain orthogonal.
Consequently,
\[
 \|\Pi W_{X,Y}|0\rangle\|^2
 =\sum_{\lambda=1}^M
   \|\psi_\lambda(X)\|^2 \chi_\lambda\omega(v_\lambda).
\]
When the function value is zero, all $\chi_\lambda$ vanish, so the accepting probability is zero.
When the function value is one, only $\ell(X)$ has a nonzero flag, and
$v_{\ell(X)}=u(X)$ with $u(X)/2^{H_D}={p_D}(X)\ge24/25>1/4$.
The same formula therefore gives
\[
 \|\Pi W_{X,Y}|0\rangle\|^2
 ={p_D}(X)\frac{2^{H_D}}{4u(X)}=\frac14.
\]
Thus, on every promised input, the accepting probability is exactly ${F_{\mathrm{par}}}(X,Y)/4$.
The circuit $W$ uses $q_D+1+s$ queries and retains all intermediate records, so its inverse has the same query cost.
Lemma~\ref{prelim:lem:exact-amplification} gives an exact algorithm using at most $3(q_D+1+s)$ queries.
\end{proof}

\subsection{Randomized Lower Bound}
\label{cp-randomized}

The probability table distinguishes the two types of input only at the true cell, so we must bound the probability that a randomized algorithm finds that cell.
We first use the Forrelation lower bound to control address prediction under the encoding promise, and then couple the two types of input in Definition~\ref{cp-function} to analyze the first visit to the true cell.

\begin{proposition}[Randomized query complexity of $F_{\mathrm{par}}$]\label{cp-lower-bound}
The partial function ${F_{\mathrm{par}}}$ in Definition~\ref{cp-function} satisfies
\begin{equation}\label{cp-general-r}
 R({F_{\mathrm{par}}})=\Omega\!\left(\min\{R(\operatorname{Forr}_{k,M})/J^2,M\}\right).
\end{equation}
\end{proposition}
\begin{proof}
We first obtain a hard distribution with prediction advantage $O(1/J)$ for a single instance.
A direct product bound, applied before conditioning on the encoding promise, then limits the probability of recovering the address.
Finally, we couple the zero and one inputs and show that distinguishing them requires visiting the true table cell.

\paragraph{A hard distribution for small advantage.}
Suppose an algorithm succeeds with probability at least $1/2+1/(100J)$ on every valid input.
Repeating it independently $r_1=\lceil5000J^2\ln3\rceil$ times and taking the majority gives error at most $e^{-r_1/(5000J^2)}\le1/3$, by Hoeffding's inequality~\cite{Hoeffding63}.
There is an absolute constant $C_1$ such that $r_1\le C_1J^2$.
Fix $0<c<1/(2C_1)$.
For each valid $X$, let $Y^1(X)$ be the promised table for which $F_{\mathrm{par}}(X,Y^1(X))=1$.
If $R(\operatorname{Forr}_{k,M})/J^2<2/c$, the claimed bound follows from $R({F_{\mathrm{par}}})\ge1$: for every valid $X$, the inputs $(X,0)$ and $(X,Y^1(X))$ have opposite function values, so an algorithm making no queries cannot have error at most $1/3$ on both.
We henceforth assume $R(\operatorname{Forr}_{k,M})/J^2\ge2/c$ and set
\[
 T=\left\lfloor cR(\operatorname{Forr}_{k,M})/J^2\right\rfloor.
\]
If $T$ queries achieved the above advantage, amplification would use at most
\[
 r_1T\le C_1cR(\operatorname{Forr}_{k,M})<R(\operatorname{Forr}_{k,M})
\]
queries, contradicting the definition of $R(\operatorname{Forr}_{k,M})$.
Hence every randomized algorithm using at most $T$ queries has a valid input on which its success probability is less than $1/2+1/(100J)$.
Yao's minimax principle (Lemma~\ref{prelim:lem:yao}) therefore gives a single distribution $\mu$ on valid inputs such that every randomized algorithm $A$ using at most $T$ queries satisfies
\[
 \Pr_{x\sim\mu,A}[A(x)=\operatorname{Forr}_{k,M}(x)]
 \le\frac12+\frac1{100J}.
\]
Fix such a distribution $\mu$ for the remainder of the proof.

Independently sample $x\sim\mu$ and a uniform bit $z\in\{0,1\}$.
This defines an input distribution $\nu$ for $g(x,z)=\operatorname{Forr}_{k,M}(x)\oplus z$, with $\nu(x,z)=\mu(x)/2$.
For every fixed $x$, the uniform bit $z$ makes the two answers equally likely, so
\[
 \Pr_{(x,z)\sim\nu}[g(x,z)=0]
 =\Pr_{(x,z)\sim\nu}[g(x,z)=1]=\frac12.
\]
Given any algorithm $B$ for $g$ using at most $T$ queries, fix $z$, answer queries to $z$ without consulting the input, and output $B(x,z)\oplus z$.
This gives an algorithm for $\operatorname{Forr}_{k,M}$ using at most $T$ queries.
Therefore
\[
 \begin{aligned}
 \Pr_\nu[B(x,z)=g(x,z)]
 &=\frac12\sum_{z\in\{0,1\}}
   \Pr_{x\sim\mu}[B(x,z)\oplus z=\operatorname{Forr}_{k,M}(x)]\\
 &\le\frac12+\frac1{100J}.
 \end{aligned}
\]

\paragraph{Conditioning on the encoding promise.}
Let $V(X)$ be the answer vector of the $J$ instances.
Under $\nu^{\otimes J}$, its coordinates are independent uniform bits.
Let $\mathcal E=\{V(X)\in\operatorname{im}(\operatorname{Enc})\}$.
By injectivity of the encoding,
\[
 \Pr[\mathcal E]=\frac M{2^J}>0.
\]
Set $\rho=\nu^{\otimes J}\mid\mathcal E$.
Consider an address predictor $\widehat\ell$ using at most $T$ queries to the base input.
If this query bound is guaranteed only on valid base inputs, truncate it after $T$ queries and output a fixed address on any unfinished branch, extending it to all inputs.
This extension agrees with the predictor on the promise domain and satisfies the same query bound under $\nu^{\otimes J}$.

Encoding the predicted address gives a prediction of all $J$ answers.
The event $\operatorname{Enc}(\widehat\ell)=V(X)$ implies $\mathcal E$, and within $\mathcal E$ it is equivalent to $\widehat\ell=\ell(X)$.
Since the algorithm makes at most $T$ queries in total, it makes at most $T$ queries to each instance.
By Lemma~\ref{lem:local-budget-product}, the probability of correctly predicting all $J$ answers under $\nu^{\otimes J}$ is at most $(1/2+1/(100J))^J$.
It follows that
\begin{align}
 \Pr_{X\sim\rho}[\widehat\ell=\ell(X)]
 &=\frac{\Pr_{\nu^{\otimes J}}[\operatorname{Enc}(\widehat\ell)=V(X)]}{\Pr[\mathcal E]}
 \notag\\
 &\le\frac{2^J(1/2+1/(100J))^J}{M}
 =\frac{(1+1/(50J))^J}{M}
 \le\frac{e^{1/50}}M.
 \label{cp-prediction}
\end{align}
The direct product bound is applied before conditioning; it does not require the instances to remain independent under $\rho$.

\paragraph{The first visit to the true cell.}
Fix an algorithm $A$ whose query complexity is $t\le T$.
Sample $X\sim\rho$ and run $A$ on $(X,0)$ and $(X,Y^1(X))$ using the same random coins.
Write $\tau_0,\tau_1$ for the resulting transcripts and $\mathcal T_0,\mathcal T_1$ for their distributions; the transcripts include the random coins and the output.

For each $j\in[t]$, simulate the algorithm for at most $t$ queries with the zero table, answering every table query by zero without consulting the input, and output the cell address of the $j$th table query.
If that query does not occur, output a fixed address.
This is an address predictor making at most $t\le T$ queries to the base input; the truncation described above ensures that the query bound also holds outside the promise domain.
Let $\mathcal H_j$ be the event that this table query occurs and visits $\ell(X)$, and let $\mathcal H=\bigcup_{j=1}^t\mathcal H_j$.
Equation~\eqref{cp-prediction} gives
\[
 \Pr[\mathcal H]
 \le\sum_{j=1}^t\Pr[\mathcal H_j]
 \le\frac{e^{1/50}t}{M}.
\]
The two tables differ only in cell $\ell(X)$.
Induction on the query steps shows that, before the first visit to that cell, the two runs receive the same answers and choose the same next query.
On $\mathcal H^c$, the complete transcripts are identical.
The coupling inequality therefore gives
\[
 \TV(\mathcal T_0,\mathcal T_1)
 \le\Pr[\tau_0\ne\tau_1]
 \le\Pr[\mathcal H]
 \le\frac{e^{1/50}t}{M}.
\]
If $A$ computes ${F_{\mathrm{par}}}$ with error at most $1/3$, the probability of output one is at least $2/3$ under $\mathcal T_1$ and at most $1/3$ under $\mathcal T_0$.
Hence
\[
 \frac13\le\TV(\mathcal T_0,\mathcal T_1)
 \quad\Longrightarrow\quad t\ge\frac{M}{3e^{1/50}}.
\]

Finally, our assumption $R(\operatorname{Forr}_{k,M})/J^2\ge2/c$ gives
\[
 T=\left\lfloor\frac{cR(\operatorname{Forr}_{k,M})}{J^2}\right\rfloor
 \ge\frac{cR(\operatorname{Forr}_{k,M})}{2J^2}.
\]
Every correct algorithm either uses more than $T$ queries or satisfies the preceding $M/(3e^{1/50})$ lower bound.
Thus
\[
 R({F_{\mathrm{par}}})\ge\min\left\{\frac{cR(\operatorname{Forr}_{k,M})}{2J^2},\frac{M}{3e^{1/50}}\right\}
 =\Omega(\min\{R(\operatorname{Forr}_{k,M})/J^2,M\}).
\]
This proves~\eqref{cp-general-r}, including the case in which $R(\operatorname{Forr}_{k,M})/J^2$ is bounded.
\end{proof}

\subsection{Separation Theorem}
\label{cp-parameters}

Combining the exact query upper bound, the randomized lower bound, and the input length in~\eqref{cp-length} gives the separation for fixed $k,d$.
We then choose these constants as a function of $\varepsilon$ to obtain constant exact quantum query complexity and nearly linear randomized query complexity.

\begin{theorem}\label{cp-main}
For every fixed $\varepsilon\in(0,1)$ and all sufficiently large $n$, there is a partial Boolean function $F_{\mathrm{par},\varepsilon,n}$ on $n$ bits satisfying
\[
 Q_E(F_{\mathrm{par},\varepsilon,n})=O_\varepsilon(1),\qquad
 R(F_{\mathrm{par},\varepsilon,n})=\Omega_\varepsilon(n^{1-\varepsilon}).
\]
\end{theorem}
\begin{proof}
We choose $k,d$ so that the quantum query complexity is constant and the randomized lower bound is $\Omega_\varepsilon(n^{1-\varepsilon})$.
After estimating the input length, we use padding to cover every sufficiently large length.

Fix $\varepsilon\in(0,1)$, choose
\[
 k=\lceil4/\varepsilon\rceil,\qquad d=\lceil12/\varepsilon\rceil,
\]
and let $b_{\mathrm{enc}}$ grow in the construction of Definition~\ref{cp-function}.
These constants satisfy $d(1-1/k)>2$.
Since $H_{\mathrm F}=O(q_{\mathrm F}db_{\mathrm{enc}})$, the number of blocks encoding the success probability satisfies
\[
 s=\left\lceil\frac{d(2H_{\mathrm F}+2b_{\mathrm{enc}})+1}{b_{\mathrm{enc}}}\right\rceil
 =O(q_{\mathrm F}d^2).
\]
Substituting this estimate and $q_D=2d(q_{\mathrm F}+1)$ into~\eqref{cp-q-cost}, and applying Corollary~\ref{cp-seed-algorithm}, gives
\begin{equation}\label{cp-q-main}
 \begin{aligned}
 Q_E(F_{\mathrm{par}})&\le3(q_D+1+s)=O(q_{\mathrm F}d^2)\\
 &=O\!\left(kd^2 2^{10k}\log(d+1)\right)=2^{O(1/\varepsilon)}.
 \end{aligned}
\end{equation}
Thus both the upper bound on exact quantum query complexity and the upper bound on $s$ are independent of $b_{\mathrm{enc}}$.

Using $J=dM^{1/d}$ and $2^{b_{\mathrm{enc}}}=M^{1/d}$, equation~\eqref{cp-length} gives
\[
 n=(dk+s)M^{1+1/d}+dM^{1/d}+M
   =\Theta_\varepsilon(M^{1+1/d}).
\]
Equation~\eqref{cp-seed-lower} yields
$R(\operatorname{Forr}_{k,M})/J^2
=\Omega_\varepsilon(M^{1-1/k-2/d}/\log M)$.
Since $0<1-1/k-2/d<1$, both terms in the minimum in~\eqref{cp-general-r} are at least this large asymptotically.
Consequently,
\begin{equation}\label{cp-r-main}
 \begin{aligned}
 R(F_{\mathrm{par}})
 &=\Omega_\varepsilon\!\left(\frac{M^{1-1/k-2/d}}{\log M}\right)\\
 &=\Omega_\varepsilon\!\left(
     \frac{n^{[d(1-1/k)-2]/(d+1)}}{\log n}\right).
 \end{aligned}
\end{equation}
Here we used $M=\Theta_\varepsilon(n^{d/(d+1)})$ and $\log M=\Theta_\varepsilon(\log n)$.
Our parameter choices satisfy
\[
 1-\frac{d(1-1/k)-2}{d+1}
 =\frac{d/k+3}{d+1}
 \le\frac1k+\frac3d\le\frac\varepsilon2.
\]
For sufficiently large $n$, we also have $\log n\le n^{\varepsilon/2}$.
Equation~\eqref{cp-r-main} therefore gives $R(F_{\mathrm{par}})=\Omega_\varepsilon(n^{1-\varepsilon})$.

Finally, the natural construction lengths are $\Theta_\varepsilon(2^{(d+1)b_{\mathrm{enc}}})$.
They tend to infinity, and the ratio of consecutive lengths has an upper bound depending only on $\varepsilon$.
For any sufficiently large target length $n$, take the term $n'$ immediately preceding the first natural length that exceeds $n$.
Then $n'\le n$ and $n'=\Theta_\varepsilon(n)$.
Padding with $n-n'$ irrelevant input bits preserves both query complexities, so the conclusion holds for all sufficiently large $n$.
\end{proof}

%% file: sections/communication_appendix.tex
\section{Communication Separations}\label{app:communication}

Replacing each input bit by an inner product gadget of logarithmic length transfers the query separations in Theorems~\ref{tot:thm:query} and~\ref{cp-main} to communication complexity.
We first prove a common conversion, then apply it to total functions and partial functions.

\subsection{From Query Complexity to Communication Complexity}\label{app:conversion}

We apply randomized lifting and exact query simulation to the same inner product gadget.

Let $n_0\ge2$, and set
\[
 \ell=\lceil 40000\log n_0\rceil,
 \qquad
 \mathrm{IP}_{\ell}(u,v)=\bigoplus_{j=1}^{\ell}u_jv_j.
\]
We use the following randomized lifting theorem for the inner product gadget.

\begin{theorem}[Lifting with inner product~{\cite[Theorems~1,14]{CFKMP19}}]
\label{thm:ip-lifting-cited}
Let $\mathcal O$ be a finite output set, and let $S\subseteq\{0,1\}^{n_0}\times\mathcal O$ be a search relation with at least one valid output for every input.
If a randomized communication protocol with public randomness solves $S\circ\mathrm{IP}_{\ell}^{n_0}$ using at most $c$ bits of communication in the worst case and with error at most $\epsilon$ on every input, then a randomized decision tree solves $S$ using $O(c/\ell)$ queries in the worst case and with error at most $\epsilon+1/10$ on every input.
\end{theorem}

For a partial function $f:\dom(f)\subseteq\{0,1\}^{n_0}\to\{0,1\}$, define
\begin{equation}\label{eq:merged-lift-definition}
 G(x,y)=f\bigl(\mathrm{IP}_{\ell}(x_1,y_1),\ldots,
                    \mathrm{IP}_{\ell}(x_{n_0},y_{n_0})\bigr),
 \qquad x_i,y_i\in\{0,1\}^{\ell}.
\end{equation}
The promise is that the $n_0$-bit string inside the parentheses belongs to $\dom(f)$.
If $f$ is total, then $G$ is also total; otherwise, $G$ retains the corresponding promise.
Each party holds $n_0\ell$ input bits, so
\begin{equation}\label{eq:merged-lift-length}
 n=2n_0\ell=\Theta(n_0\log n_0),
 \qquad \log n=\Theta(\log n_0).
\end{equation}

We use the query simulation of~\cite[Theorem~2.1 and its subsequent extension to finite alphabets]{BCW98}, with alphabet $\{0,1\}^{\ell}$ and the inner product gadget.
\begin{lemma}[Exact query simulation~{\cite[Theorem~2.1]{BCW98}}]
\label{lem:exact-query-simulation}
Let $n_0\ge2$ and $\ell\ge1$, and let $f$ be a partial Boolean function on $n_0$ input bits.
For $G=f\circ\mathrm{IP}_{\ell}^{n_0}$, we have
\[
 \Qecc(G)=O\!\left(\QE(f)(\log n_0+\ell)\right).
\]
The resulting exact protocol uses neither prior entanglement nor public randomness.
\end{lemma}

\begin{theorem}[Lifting query complexity]\label{thm:merged-lifting}
The construction above satisfies
\begin{align}
 \Qecc(G)&=O\bigl(\QE(f)\log n_0\bigr),\label{eq:merged-lift-quantum}\\
 \Rcc(G)&=\Theta\bigl(\R(f)\log n_0\bigr).
 \label{eq:merged-lift-random}
\end{align}
\end{theorem}
\begin{proof}
The exact quantum upper bound follows from coherent query simulation.
For the randomized lower bound, we express the partial function as a total search relation before applying the lifting theorem.
Simulating a randomized decision tree gives the matching upper bound.

Lemma~\ref{lem:exact-query-simulation} and $\ell=\Theta(\log n_0)$ give~\eqref{eq:merged-lift-quantum}.
For the randomized bounds, we apply Theorem~\ref{thm:ip-lifting-cited} by expressing the partial function as the search relation
\[
 S(z)=
 \begin{cases}
  \{f(z)\},&z\in\dom(f),\\
  \{0,1\},&z\notin\dom(f).
 \end{cases}
\]
Solving $S$ requires output $f(z)$ when $z\in\dom(f)$ and allows either output otherwise.
Thus, an algorithm for $S$ computes $f$ on its promise domain.
Conversely, an algorithm for $f$ truncated after its maximum number of queries on valid inputs solves $S$.
In particular, $R(S)=R(f)$; testing membership in $\dom(f)$ is unnecessary.

If the original protocol bounds its communication only on the promise domain, truncate every execution at that bound and output an arbitrary bit on any unfinished branch.
This leaves its behavior on valid inputs unchanged, and either output remains valid outside the promise.
We therefore obtain a protocol for the relation whose communication is bounded on all inputs.

Suppose that $G$ has a randomized protocol with public randomness using $c$ bits of communication with error at most $1/3$.
Repeat it independently a constant number of times and take the majority, reducing the error on valid inputs to at most $1/10$.
Either output is still valid outside the promise.
Applying Theorem~\ref{thm:ip-lifting-cited} to the resulting protocol gives a randomized decision tree making $O(c/\ell)$ queries in the worst case and with error at most $1/10+1/10=1/5<1/3$.
Restricting this tree to $\dom(f)$ gives a randomized algorithm for $f$.
Hence
\[
 R(f)=O(c/\ell),\qquad c=\Omega(\ell R(f)).
\]
For the upper bound, the parties use public random coins to simulate a randomized decision tree for $f$.
Whenever the tree queries bit $i$, Alice sends $x_i$, and Bob returns the inner product bit.
Both parties know all previous answers, so both know the next query address without communicating it.
Each simulated query uses $\ell+1$ bits of communication, giving the upper bound $O(\ell\R(f))$.
This proves~\eqref{eq:merged-lift-random}.
\end{proof}

\subsection{Total Functions}\label{app:total-communication}

We apply the conversion to the total functions in Theorem~\ref{tot:thm:query} and use their AND restriction to determine the exact quantum communication complexity.

\begin{corollary}[Total functions]\label{thm:total-communication}
Let $F_{\mathrm{tot},h}$ be the family in Theorem~\ref{tot:thm:query}, with input length $n_h$, data length $b_h$, and $t_h=\QE(F_{\mathrm{tot},h})$.
For all sufficiently large integers $h$, set $G_{\mathrm{tot},h}=F_{\mathrm{tot},h}\circ\IP_\ell^{n_h}$ with $\ell=\lceil40000\log n_h\rceil$.
Then
\[
 \Qecc(G_{\mathrm{tot},h})=\Theta(t_h\log t_h),\qquad
 \Qeccs(G_{\mathrm{tot},h})=\Theta(t_h\log t_h).
\]
Let $q$ denote either of these actual exact communication complexities.
The total input length of both parties is $q^{O(1)}$, and
\[
 \Rcc(G_{\mathrm{tot},h})\ge q^2\,2^{-O(\sqrt{\log q})}=q^{2-o(1)}.
\]
\end{corollary}

\begin{proof}
We first lift the query upper and lower bounds.
An AND restriction then yields a matching exact communication lower bound through matrix rank, including with prior entanglement.
Finally, we express the separation in terms of the actual exact communication complexity.

Theorem~\ref{tot:thm:query} gives
\begin{equation}\label{eq:merged-total-query-assumptions}
 \QE(F_{\mathrm{tot},h})=t_h,\quad n_h=t_h^{O(1)},\quad
 \R(F_{\mathrm{tot},h})\ge t_h^2\,2^{-O(\sqrt{\log t_h})}.
\end{equation}
Moreover, $F_{\mathrm{tot},h}$ has an $\AND_{b_h}$ restriction with $b_h=\Theta(t_h)$.
Since $b_h=\Theta(t_h)$ and $b_h\le n_h=t_h^{O(1)}$, we have
\[
 \log n_h\ge\log t_h-O(1),\qquad \log n_h=O(\log t_h).
\]
Thus $\log n_h=\Theta(\log t_h)$ and $\ell=\Theta(\log t_h)$.
Theorem~\ref{thm:merged-lifting} yields
\begin{equation}\label{eq:merged-total-cc-upper}
 \Qecc(G_{\mathrm{tot},h})=O(t_h\log t_h).
\end{equation}
The same theorem gives the randomized bound
\[
 \Rcc(G_{\mathrm{tot},h})=\Theta(\R(F_{\mathrm{tot},h})\log t_h).
\]
Combining this equality with the randomized query lower bound in~\eqref{eq:merged-total-query-assumptions} gives
\begin{equation}\label{eq:merged-total-cc-random}
 \Rcc(G_{\mathrm{tot},h})\ge t_h^2\log t_h\,2^{-O(\sqrt{\log t_h})}.
\end{equation}

The $\mathrm{AND}_{b_h}\circ\mathrm{IP}_{\ell}$ restriction gives a matching exact quantum lower bound.
For each outer coordinate fixed to $\sigma\in\{0,1\}$ in the $\AND_{b_h}$ restriction, fix the corresponding inner product input pair to $(e_1,\sigma e_1)$, where $e_1$ is the first standard basis vector of $\mathbb F_2^\ell$.
This input pair has inner product $\sigma$.
Leave unrestricted the input pairs corresponding to the $b_h$ free outer bits.
The resulting communication function is exactly
\[
 ((u_1,\ldots,u_{b_h}),(v_1,\ldots,v_{b_h}))
 \longmapsto\prod_{i=1}^{b_h}\IP_\ell(u_i,v_i).
\]
Fixing local input bits cannot increase either exact quantum communication complexity.

Let $K_\ell$ be the $0/1$ communication matrix of inner product, let $W_\ell=((-1)^{x\cdot y})_{x,y}$ be the unnormalized Walsh--Hadamard matrix, and let $\mathbf J_\ell$ be the matrix of the same size with every entry equal to one.
The identity $W_\ell W_\ell^{\mathsf T}=2^\ell I$ shows that $W_\ell$ is invertible, and
\[
 K_\ell=\frac{\mathbf J_\ell-W_\ell}{2},\qquad
 \rank K_\ell\ge\rank W_\ell-\rank\mathbf J_\ell=2^\ell-1.
\]
The row of $K_\ell$ indexed by the zero vector is zero, so its rank is exactly $2^\ell-1$.
Conjunction is multiplication in the $0/1$ representation.
Hence the communication matrix of the restriction above is $K_\ell^{\otimes b_h}$, and
\[
 \rank(K_\ell^{\otimes b_h})=(2^\ell-1)^{b_h}.
\]
The exact quantum log-rank lower bound, which allows arbitrary prior entanglement~\cite{BW01,MW07}, gives an absolute constant $c>0$ such that
\begin{equation}\label{eq:merged-total-cc-lower}
 \Qeccs(G_{\mathrm{tot},h})
 \ge c\log\rank(K_\ell^{\otimes b_h})
 =c b_h\log(2^\ell-1).
\end{equation}
We use only the asymptotic form of this bound; the formulation involving $\rank M-1$ in~\cite[Theorem~7.2.3]{dW01} gives the same conclusion.
Since $\log(2^\ell-1)\ge\ell/2$ for $\ell\ge2$, $b_h=\Theta(t_h)$, and $\ell=\Theta(\log t_h)$, it follows that
\[
 \Qeccs(G_{\mathrm{tot},h})=\Omega(t_h\log t_h).
\]

Allowing prior entanglement cannot increase communication complexity.
Equations~\eqref{eq:merged-total-cc-upper} and~\eqref{eq:merged-total-cc-lower} therefore imply
\[
 \Qeccs(G_{\mathrm{tot},h})=\Theta(t_h\log t_h),
 \qquad \Qecc(G_{\mathrm{tot},h})=\Theta(t_h\log t_h).
\]
Let $q$ denote either of these actual exact communication complexities.
The matching bounds give
\[
 \log q=\log t_h+\log\log t_h+O(1)=\Theta(\log t_h).
\]
Thus $t_h=\Theta(q/\log t_h)=\Theta(q/\log q)$, and
\[
 t_h^2\log t_h=\Theta(q^2/\log q).
\]
Substituting into~\eqref{eq:merged-total-cc-random} yields
\begin{equation}\label{eq:merged-total-final}
 \Rcc(G_{\mathrm{tot},h})\ge
 \frac{q^2}{\log q}\,2^{-O(\sqrt{\log q})}
 =q^2\,2^{-O(\sqrt{\log q})}.
\end{equation}
The last expression uses $1/\log q=2^{-\log\log q}$ and $\log\log q=O(\sqrt{\log q})$ to absorb the additional logarithmic loss into the subpolynomial factor.
The total input length of both parties is $n=2n_h\ell=t_h^{O(1)}=q^{O(1)}$.
\end{proof}

\subsection{Partial Functions}\label{app:partial-communication}

Applying Theorem~\ref{thm:merged-lifting} to the functions in Theorem~\ref{cp-main} and padding with unused input bits gives the following separation.

\begin{corollary}[Partial functions]\label{cor:partial-communication}
For every fixed $\varepsilon\in(0,1)$ and every sufficiently large $n$, there is a partial communication function $G_{\mathrm{par},\varepsilon,n}$ with total input length exactly $n$ such that
\[
 \Qecc(G_{\mathrm{par},\varepsilon,n})=O_{\varepsilon}(\log n),\qquad
 \Rcc(G_{\mathrm{par},\varepsilon,n})
 =\Omega_{\varepsilon}\bigl(n^{1-\varepsilon}(\log n)^{\varepsilon}\bigr).
\]
The quantum upper bound is achieved by an exact protocol without prior entanglement.
The implicit constants may depend on the fixed $\varepsilon$.
\end{corollary}

\begin{proof}
We choose the original query length so that its lift fits within the target communication input length.
Applying the conversion theorem and padding with unused bits then gives the stated bounds.

Fix $\varepsilon\in(0,1)$ and a sufficiently large target length $n$, and set
\[
 n_0=\left\lfloor\frac{n}{160000\log n}\right\rfloor,
 \qquad \ell=\lceil40000\log n_0\rceil.
\]
Then $n_0=\Theta(n/\log n)$ and $\ell=\Theta(\log n_0)=\Theta(\log n)$.
By Theorem~\ref{cp-main}, there is a partial function $F_{\mathrm{par}}$ on exactly $n_0$ input bits satisfying
\[
 \QE(F_{\mathrm{par}})=O_\varepsilon(1),\qquad
 \R(F_{\mathrm{par}})=\Omega_\varepsilon(n_0^{1-\varepsilon}).
\]
Its inner product lift has total input length $2n_0\ell$.
Since $n_0\le n$,
\[
 2n_0\ell
 \le\frac{2n}{160000\log n}(40000\log n+1)
 =\frac n2+\frac{n}{80000\log n}\le n.
\]
Add a total of $n-2n_0\ell$ input bits to the two parties, with arbitrary values and ignored by the function, to obtain $G_{\mathrm{par},\varepsilon,n}$.
The original protocols can ignore the additional bits; conversely, fixing them to zero recovers the lifted function.
Thus padding preserves both communication complexities.

Applying Theorem~\ref{thm:merged-lifting} to $F_{\mathrm{par}}\circ\IP_\ell^{n_0}$ gives
\[
 \Qecc(G_{\mathrm{par},\varepsilon,n})
 =O_\varepsilon(\log n_0)=O_\varepsilon(\log n)
\]
and
\begin{equation}\label{eq:merged-partial-near-linear}
 \begin{aligned}
 \Rcc(G_{\mathrm{par},\varepsilon,n})
 &=\Omega_\varepsilon(n_0^{1-\varepsilon}\log n_0)\\
 &=\Omega_\varepsilon\!\left(
     \left(\frac n{\log n}\right)^{1-\varepsilon}\log n\right)\\
 &=\Omega_\varepsilon\bigl(n^{1-\varepsilon}(\log n)^\varepsilon\bigr).
 \end{aligned}
\end{equation}
Together with the quantum upper bound, this proves the corollary.
The promise is that the string of inner product outputs belongs to the promise domain of the original query function, so the resulting function remains partial.
\end{proof}

%% file: sections/acknowledgments.tex
\section*{Acknowledgments}
This work was supported by the research startup fund of Fuzhou University (Grant No.\ \mbox{XRC-26004}).

%% file: sections/ai_methodology.tex
\section*{AI Methodology Statement}\label{sec:ai-methodology}
The author used ChatGPT (OpenAI) to assist with developing and organizing the constructions and proof arguments in Sections~\ref{sec:total}--\ref{cp-section} and Appendix~\ref{app:communication}.
The author is responsible for the correctness, originality, and accuracy of the manuscript, including its references.